\documentclass[letterpaper,journal]{IEEEtran}
\usepackage{amsmath,amsfonts,amssymb,amsthm}
\usepackage{algorithmic}
\usepackage{array}
\usepackage{subcaption}
\usepackage{textcomp}
\usepackage{cite}
\usepackage{stfloats}
\usepackage{url}
\usepackage{verbatim}
\usepackage{graphicx}
\usepackage{enumitem}
\usepackage{algorithm}
\usepackage{color}
\usepackage{xcolor}
\usepackage{comment}

\newtheorem{remark}{Remark}
\theoremstyle{definition} 
\newtheorem{assumption}{Assumption}
\def\BibTeX{{\rm B\kern-.05em{\sc i\kern-.025em b}\kern-.08em
    T\kern-.1667em\lower.7ex\hbox{m_r}\kern-.125emX}}
\usepackage{balance}

\begin{document}
\title{Joint Communication and Sensing Security in MIMO-OFDM ISAC Systems via Artificial Noise and Ghosts}
\author{Yinchao Yang, Prabhat Raj Gautam, Yathreb Bouazizi,  Mohammad Kazemi, Michael Breza, and Julie A. McCann
\thanks{The authors are with the Department of Computing, Imperial College London, U.K., SW7 2AZ. (e-mails: \{yinchao.yang; p.gautam; y.bouazizi18; mohammad.kazemi; michael.breza04; j.mccann\}@imperial.ac.uk).}
\thanks{This work was supported by the EPSRC and DSIT through the Communications Hub for Empowering Distributed Cloud Computing Applications and Research (CHEDDAR) [grant numbers EP/X040518/1 and EP/Y037421/1].} 
}

\markboth{Journal of \LaTeX\ Class Files,~Vol.~18, No.~9, September~2020}%
{How to Use the IEEEtran \LaTeX \ Templates}

\maketitle

\begin{abstract}
Integrated sensing and communication (ISAC) enables efficient sharing of wireless resources to support emerging applications, but it also introduces new sensing-based security vulnerabilities. Here, potential communication security threats include confidential messages intended for legitimate users being intercepted, and unauthorized receivers (Eves) passively exploiting target echoes to infer sensing parameters without users being aware. Despite these risks, the joint protection of sensing and communication security in ISAC systems remains unexplored. To address this challenge, this paper proposes a dual-secure ISAC framework that simultaneously protects sensing and communication against passive sensing Eves and communication Eves, without requiring their channel state information (CSI). Specifically, transmit beamformers are jointly designed to inject artificial noise (AN) to introduce interference to communication Eves, while deliberately distorting the reference signal available to sensing Eves to impair their sensing capability. Furthermore, the proposed design generates artificial ghosts (AGs) with fake angle-delay-Doppler profiles. The legitimate sensing receiver can suppress these AGs using its prior knowledge of the true target directions, whereas sensing Eves cannot, thereby significantly reducing their probability of correctly detecting the true targets. Numerical results demonstrate that the proposed framework effectively enhances both communication and sensing security, while preserving the performance of communication users and legitimate sensing receivers.

\end{abstract}


\section{Introduction}

Integrated sensing and communication (ISAC) has been recognized by the International Telecommunication Union (ITU) as one of the key enabling technologies for sixth-generation (6G) wireless systems. By unifying sensing and communication on a common spectrum and hardware, ISAC enables simultaneous transmission of information and perception of the surrounding environment \cite{liu2022integrated}. More importantly, the ISAC framework makes sensing and communication mutually beneficial. On one hand, wireless sensing provides real-time environmental awareness, such as object locations and mobility states, which can be exploited by communication systems to perform adaptive beamforming, handover optimization, and interference management. On the other hand, communication enables sensing systems to exchange information for cooperative perception and network-level data fusion \cite{meng2025cooperative}. Millimeter-wave (mmWave) ISAC, in particular, enables a wide range of emerging applications, such as localization in autonomous vehicular networks \cite{zhen2025towards} and vital-sign detection \cite{yang2025toward}.

However, the reuse of the signal for both sensing and communication makes physical-layer security concerns more critical in ISAC systems, as it exposes both functionalities to potential security threats.
The first security concern arises in the communication domain, where an eavesdropper (Eve) may intercept the transmitted waveform and attempt to recover confidential information if the encryption key is known or compromised. For instance, the target being detected by the transmitter can be the communication Eve, who receives the joint signal and decodes confidential information, leading to data security breaches. The second and more distinctive threat lies in sensing security. Owing to the broadcast nature of electromagnetic (EM) waves, a sensing Eve can receive reflected echoes from targets while simultaneously overhearing the direct-path signal from the transmitter as a reference waveform. By jointly processing the two signals, such as using a matched filter (MF), the sensing Eve can infer sensitive target information like velocity, geometric features, and even physiological information (e.g., respiration and heartbeat), thereby posing severe privacy risks. 

Communication security in ISAC systems, also referred to as data or information security, has attracted significant research attention in recent years, resulting in a rich body of literature that establishes both solid theoretical foundations \cite{zhu2024enabling} and practical design methodologies \cite{luo2025isac}. In particular, a variety of physical-layer keyless security techniques have been developed to safeguard confidential information against passive eavesdropping. Representative approaches include information theoretic secure designs \cite{su2023sensing, liu2025sensing}, artificial noise (AN) designs \cite{su2020secure, jia2025secure}, secure beamforming optimizations \cite{chu2022joint, xia2025towards}, and waveform design methods \cite{zhou2022integrated, zhang2024intelligent}. For active attacks, such as jamming and spoofing, in which an adversary deliberately injects interfering or deceptive signals to disrupt reliable reception or induce false information decoding, effective design methods have also been studied, e.g., \cite{pirayesh2022jamming, liu2024game}.

 
Conversely, sensing security, particularly sensing eavesdropping, is not yet mature in existing ISAC systems \cite{geng2025survey, han2025next}. Active sensing attacks, such as jamming, can often be mitigated using defensive mechanisms developed for active communication attacks \cite{cao2025sensing, li2025win}. In contrast, a passive sensing Eve does not transmit any signals, yet it can still infer sensitive physical-world information, thereby posing a fundamentally different and more challenging security threat. Existing defenses designed against passive communication Eves cannot be readily extrapolated to sensing Eves. Moreover, unlike sensing-only systems, in which secure designs based on radar-centric waveforms have been extensively investigated \cite{hernandez2023scheduled, meng2025securfi}, many ISAC frameworks employ communication-centric waveforms to simultaneously support both sensing and data transmission. Such communication-centric signaling is inherently more vulnerable to sensing eavesdroppers due to the fundamental trade-off between sensing and communication under limited spectral and power resources \cite{guo2024secure}. 
Radar-centric waveforms, such as FMCW, typically operate over wide bandwidths and exhibit low probability of interception (LPI) characteristics, while often requiring dedicated sensing hardware. Communication waveforms, however, are designed for interoperability across standard communication devices and are therefore more easily accessible to unauthorized receivers. Consequently, how to effectively prevent sensing eavesdropping in ISAC systems employing communication-centric waveforms remains largely unexplored \cite{qu2023privacy, aman2025integrating}.


To address this challenge, several recent studies have investigated sensing security against sensing Eves using various sensing metrics. In \cite{jia2024illegal}, illegal sensing is mitigated by minimizing the signal-to-noise ratio (SNR) at the sensing adversary while simultaneously minimizing the Cramér-Rao bound (CRB) at the legitimate sensing receiver to preserve sensing accuracy. In \cite{zou2024securing}, a mutual-information (MI) based framework is developed, where the MI of the legitimate sensing receiver is maximized while maintaining a maximum allowable MI at the sensing Eves and a minimum signal-to-interference-plus-noise ratio (SINR) requirements of communication users to guarantee their performance. Similarly, \cite{zhou2025beamforming} maximizes the SNR of the legitimate sensing receiver to ensure sensing performance, while imposing SINR constraints for communication users and an upper bound on the sensing Eve's SNR to achieve sensing security. A dual-secure design for cell-free ISAC networks is studied in \cite{ren2024secure}, where the sensing Eve is assumed to have no access to the reference signal and therefore can only perform non-coherent energy detection. 
Under such an assumption, \cite{ren2024secure} optimized the detection probability for the sensing Eves and minimized the communication Eve's SNR to ensure communication security. The above-mentioned works rely on the common assumption that the channel state information (CSI) of sensing Eves is perfectly known at the transmitter, which would make these existing designs optimistic 
and impractical in realistic deployments.

For the case where the CSI of sensing Eves is unknown, i.e., the sensing Eve-agnostic scenario, the authors in \cite{musallam2025enhancing} enhance sensing security by minimizing the signal-to-artificial-noise ratio while maximizing the SNR of communication users to guarantee communication quality. However, data security against communication Eves and the sensing performance of the legitimate sensing receiver are not considered. In \cite{han2025sensing}, artificial peaks are created via the ambiguity function (AF) shaping to control subcarriers' power allocation in OFDM systems, resulting in the coexistence of actual and ghost targets in the sensing Eve's range profile. However, having the ghost targets confined to the range domain rather than the full angle–range–Doppler domain makes it easier for sensing Eves to correctly identify the true targets. In a similar vein, \cite{du2025securing} investigates both OTFS and OFDM waveforms for sensing security by optimizing a Kullback–Leibler divergence–based detection metric, such that strong sidelobes are induced in the sensing Eve's range–Doppler profile to obscure the mainlobe. However, both works assume that the sensing Eve's reference signal contains non-line-of-sight (NLoS) components that inherently degrade its quality. In practical scenarios, the sensing Eve may not always experience strong NLoS distortion. In \cite{chen2025sensing}, sensing security in near-field ISAC systems is achieved by exploiting environmental scatter to mislead sensing eavesdroppers. This approach relies on prior knowledge of the scatter information at the transmitter and assumes static scattering environments. Yet, these recent works focus solely on sensing security and do not explicitly consider communication security. An ISAC design considering dual security introduces further challenges and has not been addressed.

To bridge this research gap, where communication and sensing security must be jointly addressed in the absence of CSI of both communication and sensing Eves, we propose a dual-secure design against communication and sensing Eves, without the knowledge of their CSI, while guaranteeing the communication performance and sensing performance for communication users and legitimate sensing receivers, respectively. Our main contributions are summarized below:
\begin{enumerate}
    \item We propose a secure MIMO-OFDM ISAC design that mitigates both sensing and communication eavesdropping without requiring the CSI of potential Eves while guaranteeing the performance of legitimate users. By injecting AN, the received signals at communication Eves are deliberately degraded, thereby reducing information leakage. Meanwhile, the AN also contaminates the reference signals available to sensing Eves, introducing impairments during the MF process and limiting their estimation capability. This constitutes the first layer of protection for both communication and sensing security. 
    
    \item Even with this first layer of protection, advanced sensing Eves may be able to suppress the injected AN, allowing them to retain high-quality reference signals. To address this more capable adversarial scenario, a second layer of sensing protection is introduced by deliberately generating artificial ghosts (AGs) with fake angle-delay-Doppler profiles. Specifically, the AF is shaped to create artificial peaks in the angle-delay-Doppler domain, thereby misleading sensing Eves toward ghost targets and reducing their probability of correctly detecting the true targets. By jointly employing the two defense mechanisms, the proposed design provides sensing protection against sensing Eves with varying capabilities while simultaneously guaranteeing communication performance and security.
    
    \item We establish performance measures for communication performance, communication security, sensing performance, and sensing security, respectively. With the established performance metrics, we formulate an optimization problem that aims to maximize the sensing performance of the legitimate sensing receiver while maintaining 
    communication performance, communication security, and sensing security, under limited resources. To tackle the resulting non-convex problem, several convex surrogate functions are introduced to obtain a tractable formulation.
\end{enumerate}

The remainder of this paper is organized as follows. Section~II introduces the system model, including the underlying assumptions, the signal models of the transmitter and various receivers, and the associated signal processing procedures. Section~III presents the performance metrics adopted in the design. Section~IV formulates the proposed optimization problem and describes the solution. Section~V provides numerical results to evaluate the performance of the proposed design. Finally, Section~VI concludes the paper.

\subsection*{List of Notations:}
Capital boldface letters (e.g., $\mathbf{A}$) denote matrices, while lowercase boldface letters (e.g., $\mathbf{a}$) denote vectors. Scalars are represented using regular lowercase or uppercase fonts. 
The symbols $\mathbb{C}$, $\mathbb{C}^{n\times1}$, and $\mathbb{C}^{m\times n}$ denote the set of complex numbers, the set of $n$-dimensional complex vectors, and the set of $m\times n$ complex matrices, respectively.
The identity matrix is denoted by $\mathbf{I}$, and the zero matrix by $\mathbf{0}$. The notations $[\cdot]^{\mathrm{H}}$, $[\cdot]^{*}$, $\mathrm{Tr}(\cdot)$, $\mathrm{rank}(\cdot)$, and $\mathrm{blkdiag}(\cdot)$ represent the Hermitian transpose, the complex conjugate, the trace, the rank, and the block diagonal of a matrix, respectively, and the $\ell_2$ norm is denoted by $\left\|\cdot\right\|$. The symbol $\succeq$ indicates positive semidefinite, while $\odot$ and $\otimes$ denote element-wise and Kronecker product, respectively. Finally, $\mathcal{CN}(0, \sigma^2)$ represents a standard complex Gaussian distribution with zero mean and variance $\sigma^2$.

\section{System Model}

\begin{figure}[t]
    \centering
    \includegraphics[width=0.9\linewidth]{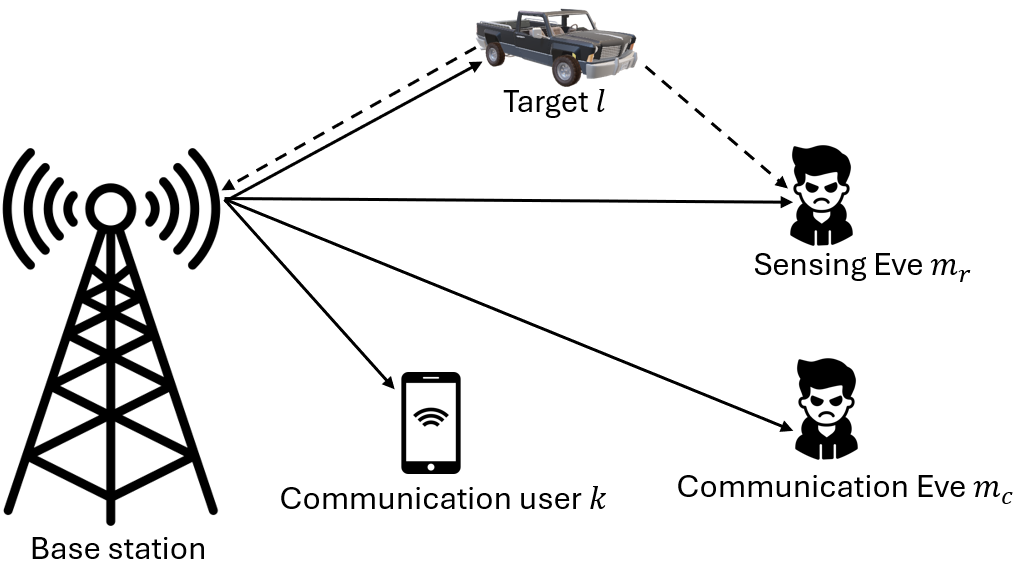}
    \caption{A monostatic ISAC system with a target $l \in \mathcal{L}$, a communication user $k \in \mathcal{K}$, a bistatic sensing eavesdropper $m_r \in \mathcal{M}_r$, and a communication eavesdropper $m_c \in \mathcal{M}_c$.}
    \label{fig:monostatic ISAC}
\end{figure}

As illustrated in Fig.~\ref{fig:monostatic ISAC}, we consider a monostatic mmWave ISAC base station (BS) equipped with $N_t$ transmit antennas and $N_r$ receive antennas, serving $K$ communication users (CUs) each equipped with $N_k$ antennas. The set of CUs is denoted by $\mathcal{K} \triangleq \{1,2,\ldots,K\}$. In addition to communication, the BS performs radar sensing to localize $L$ far-field, point-like targets, whose indices form the set $\mathcal{L} \triangleq \{1,2,\ldots, L\}$. In parallel, we consider the presence of sensing Eves equipped with $N_{m_r}$ antennas that act as bistatic receivers attempting to intercept the target parameters from reflected echoes. The sensing Eve's index set is denoted by $\mathcal{M}_r \triangleq \{1,2,\ldots,M_r\}$. Likewise, a group of communication Eves attempts to overhear the downlink communication signals. Their index set is given by $\mathcal{M}_c \triangleq \{1,2,\ldots,M_c\}$. 
Throughout this paper, we make the following assumptions:

\begin{assumption}\label{assumption1}
The sensing Eves can perfectly separate the reference signal from the surveillance signal, and they know the physical location of the BS.
\end{assumption}

\begin{assumption}\label{assumption2}
The BS has no knowledge of sensing or communication Eve's position or its existence.
\end{assumption}

\begin{assumption}\label{assumption3}
The BS knows the true angle of departure/arrival (AoD/AoA) of the targets, whereas sensing Eves do not.
\end{assumption}

\begin{remark}
Assumption~\ref{assumption1} intentionally adopts a conservative attacker model in which the sensing Eve can perfectly separate the reference and surveillance signals and knows the BS location. This gives the sensing Eve favorable conditions for recovering the transmitted reference waveform. In practice, imperfect signal separation, finite array aperture, and residual interference would further degrade its sensing capability. Assumption~\ref{assumption2} reflects fully passive Eves whose positions and CSI are unavailable to the BS. Assumption~\ref{assumption3} can be realized through an initial beam-sweeping or omnidirectional transmission stage, enabling the BS to determine the target directions before the subsequent sensing and beamforming design.
\end{remark}


\subsection{Transmit Signal Model}

The ISAC BS employs an OFDM waveform to simultaneously serve CUs and illuminate the environment for sensing. Let $\mathcal{N}_c$ and $\mathcal{N}_s$ denote the sets of subcarrier and OFDM symbol indices, respectively. For each subcarrier-symbol pair $(n_c,n_s)$, the modulated symbol vector is $\mathbf{s}_{n_c,n_s} = \left[ s_{n_c,n_s,1},\ldots,s_{n_c,n_s,K} \right]^{\mathrm{T}} \in \mathbb{C}^{K \times 1}$, where $s_{n_c,n_s,k}$ denotes the modulated symbol of user~$k$, drawn from a constellation set. The resulting transmit signal is given by
\begin{equation}\label{Tx signal}
    \mathbf{x}_{n_c,n_s} = \mathbf{W}_{n_c,n_s}\mathbf{s}_{n_c,n_s} + \mathbf{q}_{n_c,n_s},
\end{equation}
    where $\mathbf{W}_{n_c,n_s} \in \mathbb{C}^{N_t\times K}$ is the beamforming matrix with each column $\mathbf{w}_{n_c, n_s, k} \in \mathbb{C}^{N_t\times 1}$ being the beamforming vector of each user~$k$, and $\mathbf{q}_{n_c,n_s} \in \mathbb{C}^{N_t \times 1}$ is the AN vector. Following from \cite{li2024framework, liao2025design}, we assume $\mathbb{E} \left[ \mathbf{s}_{n_c,n_s} \mathbf{s}_{n_c,n_s}^{\mathrm{H}} \right] = \mathbf{I}$ and $\mathbb{E} \left[ \mathbf{s}_{n_c,n_s} \mathbf{q}_{n_c,n_s}^{\mathrm{H}} \right] = \mathbf{0}$, which indicates that the symbols are uncorrelated across users and are independent of the AN, respectively. Lastly, the AN vector is designed to lie in the null space of the target directions based on their known AoDs, as specified in Assumption~\ref{assumption3}.

The corresponding baseband OFDM signal is given by \cite{sturm2011waveform}
\begin{equation}\label{OFDM_Tx}
    \mathbf{x}(t)
    = \sum_{n_c\in\mathcal{N}_c} \sum_{n_s\in\mathcal{N}_s} \mathbf{x}_{n_c,n_s} e^{j2\pi n_c\Delta f t_{n_s}}
      \operatorname{rect}\left(\frac{t_{n_s}}{T_{\mathrm{tot}}}\right),
\end{equation}
where $t_{n_s} = t - n_s T_{\mathrm{tot}}$, $\Delta f = 1/T$ is the subcarrier spacing, $T_{\mathrm{tot}} = T + T_{\mathrm{cp}}$ denotes the total symbol duration, with $T$ and $T_{\mathrm{cp}}$ being the symbol duration and the cyclic prefix (CP) duration, respectively.

Finally, the baseband signal in \eqref{OFDM_Tx} is up-converted to the carrier frequency $f_c$, resulting in the radio frequency (RF) domain transmit waveform $\mathbf{x}_{\mathrm{RF}}(t) = \mathbf{x}(t)e^{j2\pi f_c t}$.

\subsection{Monostatic Sensing Signal Model}

The monostatic BS observes echoes reflected by all targets in the set $\mathcal{L}$. The received time-domain signal after down-conversion can be expressed as 
\begin{align}\label{multi-target echo}
    &\mathbf{y}_{B}(t)
        = \sum_{l\in\mathcal{L}}
           \tilde{\beta}_{l,B}
           \mathbf{a}_r(\theta_{l,B})
           \mathbf{a}_t^{\mathrm{H}}(\theta_{l,B})
           \sum_{n_c\in\mathcal{N}_c}
           \sum_{n_s\in\mathcal{N}_s}
           \mathbf{x}_{n_c,n_s} \\
    & \quad \cdot e^{j2\pi n_c\Delta f (t_{n_s}-\tau_{l,B})} e^{j2\pi\mu_{l,B} t}
           \operatorname{rect}\left(\frac{t_{n_s}-\tau_{l,B}}{T_{\mathrm{tot}}}\right) + {\mathbf{n}}(t), \notag
\end{align}
where $\tilde{\beta}_{l,B} = \beta_{l,B}e^{-j2\pi f_c \tau_{l,B}}$, and $\beta_{l,B}$ denotes the complex channel coefficient that incorporates propagation loss, antenna gains, and the radar cross section (RCS) of the $l$-th target. The vectors $\mathbf{a}_t(\theta_{l,B})\in\mathbb{C}^{N_t \times 1}$ and $\mathbf{a}_r(\theta_{l,B})\in\mathbb{C}^{N_{r} \times 1}$ are the transmit and receive steering vectors evaluated at the AoD/AoA. The round-trip delay and Doppler shift are denoted by $\tau_{l,B}$ and $\mu_{l,B}$, respectively. Finally, $\mathbf{n}(t)\sim\mathcal{CN}(\mathbf{0},\sigma_r^2\mathbf{I}_{N_r})$ denotes the additive white Gaussian noise (AWGN).

Following the standard OFDM signal processing procedures \cite{dai2025tutorial, li2024mimo}, we can obtain the frequency-domain signal:
\begin{align}\label{freq-domain-echo}
    \mathbf{y}_{n_c,n_s,B} = &\sum_{l\in\mathcal{L}}
        \tilde{\beta}_{l,B}
        \mathbf{a}_r(\theta_{l,B})
        \mathbf{a}_t^{\mathrm{H}}(\theta_{l,B}) \\
    & \cdot \mathbf{x}_{n_c,n_s} e^{-j2\pi n_c\Delta f\tau_{l,B}}
        e^{j2\pi \mu_{l,B} n_s T_{\mathrm{tot}}}
        + \tilde{\mathbf{n}}_{n_c, n_s}, \notag
\end{align}
where $\tilde{\mathbf{n}}_{n_c, n_s}$ is the frequency-domain AWGN.

Since the BS has prior knowledge of the AoAs/AoDs, as stated in Assumption \ref{assumption3}, the AN vector can be designed to lie in the null space of the target directions. Moreover, the receive steering vector can be eliminated, yielding the following expressions:
\begin{align} \label{estimation-signal}
        y_{n_c,n_s,B} &= \sum_{l\in\mathcal{L}}
           \tilde{\beta}_{l,B}
           \mathbf{a}_t^{\mathrm{H}}(\theta_{l,B})
           \mathbf{W}_{n_c,n_s} \mathbf{s}_{n_c, n_s} \\
        &\quad \cdot
           e^{-j2\pi n_c\Delta f\tau_{l,B}}
           e^{j2\pi\mu_{l,B} n_s T_{\mathrm{tot}}}
           + \tilde{n}_{n_c, n_s}, \notag
\end{align}
which contains target AoD, delay, and Doppler information.

\subsection{Bistatic Sensing Signal Model}

The bistatic sensing Eve receives echo signals reflected from the targets through a bistatic propagation path. The down-converted time-domain signal is modeled as
\begin{equation}
    \mathbf{y}_{m_r}(t) = \sum_{l\in\mathcal{L}} \mathbf{H}_{l,m_r} \mathbf{x}(t-\tau_{l,m_r}) e^{j2\pi \mu_{l,m_r} t} + \mathbf{n}(t),
\end{equation}
where $ \mathbf{H}_{l,m_r} = \tilde{\beta}_{l,m_r} \mathbf{a}_r(\theta_{l,m_r}) \mathbf{a}_t^{\mathrm{H}}(\theta_{l,B})$ with $\tilde{\beta}_{l,m_r} = \beta_{l,m_r} e^{-j2\pi f_c \tau_{l, m_r}}$. Here, $\beta_{l,m_r}$ denotes the bistatic channel coefficient, $\theta_{l,m_r}$ represents the AoA and $\mathbf{a}_r(\theta_{l,m_r})\in\mathbb{C}^{N_{m_r} \times 1}$ is the receive steering vector, $\tau_{l,m_r}$ and $\mu_{l,m_r}$ denote the delay and Doppler shift, respectively, and $\mathbf{n}(t)$ is AWGN.

\subsubsection{Reference Signal}
Following from Assumption~\ref{assumption1}, the sensing Eve is able to observe a reference signal directly from the BS. For a hypothetical Eve located at angle $\theta_{m_r} \in \left[-\frac{\pi}{2}, \frac{\pi}{2}\right]$, $m_{r} \notin \mathcal{K} \cup \mathcal{L}$, this observed signal in the frequency domain is denoted by
\begin{equation}\label{Eve ref signal}
    \mathbf{z}_{n_c, n_s, m_r}
    = \mathbf{J}_{n_c, n_s, m_r} \mathbf{x}_{n_c, n_s}
      + \mathbf{n}_{n_c, n_s, m_r},
\end{equation}
where $\mathbf{J}_{n_c, n_s, m_r} \in \mathbb{C}^{N_{m_r} \times N_t} = \beta_{m_r} \mathbf{a}_{r}\left(\theta_{m_r}\right) \mathbf{a}_{t}^{\mathrm{H}}\left(\theta_{m_r}\right)$ denotes the channel between BS and the sensing Eve, and
$\mathbf{n}_{n_c,n_s,m_r}$ denotes the corresponding AWGN. Since the sensing Eve is aware of the transmitter's physical location, the reference signal can be recovered via \cite{kay1993fundamentals}:
\begin{equation}\label{Eve ref signal covered}
    \hat{\mathbf{x}}_{n_c, n_s} = \left( \mathbf{J}_{n_c, n_s, m_r} \right)^{\dagger} \mathbf{z}_{n_c, n_s, m_r}, 
\end{equation}
where $(\cdot)^{\dagger}$ is the Moore–Penrose inverse.

\subsubsection{Echo Signal}
By applying the standard OFDM processing procedures, the sensing Eve obtains the following frequency-domain echo:
\begin{align} 
    \mathbf{y}_{n_c,n_s,m_r} &= \sum_{l\in\mathcal{L}}
           \tilde{\beta}_{l,m_r} \mathbf{a}_{r}\left( \theta_{l,m_r} \right) \mathbf{a}_{t}^{\mathrm{H}}\left( \theta_{l,B} \right)
           \mathbf{W}_{n_c,n_s} \mathbf{s}_{n_c,n_s} \notag \\
    &\; \cdot  e^{-j2\pi n_c\Delta f\tau_{l,m_r}}
           e^{j2\pi \mu_{l,m_r} n_s T_{\mathrm{tot}}}
           + \tilde{\mathbf{n}}_{n_c, n_s, m_r}, \label{eq9}
\end{align}
where $\tilde{\mathbf{n}}_{n_c,n_s,m_r}$ denotes frequency-domain AWGN. Since the AN vector is nulled toward the target directions, it does not appear in \eqref{eq9}.

Considering a worst-case scenario in which the sensing Eve can perfectly estimate target AoAs, the resulting echo signal becomes
\begin{align} \label{Eve processed echo no AoA}
    y_{n_c,n_s,m_r} & = \sum_{l\in\mathcal{L}}
           \tilde{\beta}_{l,m_r} \mathbf{a}_{t}^{\mathrm{H}}\left( \theta_{l,B} \right)
           \mathbf{W}_{n_c,n_s} \mathbf{s}_{n_c,n_s}\\
    &\quad \cdot  e^{-j2\pi n_c\Delta f\tau_{l,m_r}}
           e^{j2\pi \mu_{l,m_r} n_s T_{\mathrm{tot}}}
           + \tilde{n}_{n_c, n_s, m_r}, \notag
\end{align}
where the receive steering vector is effectively canceled.

\subsection{Sensing Signal Processing}

The target AoAs at the receivers are first obtained through
receive array processing. The subsequent sensing processing focuses on estimating the delay and Doppler parameters at all receivers, while sensing Eve $m_r$ additionally estimates the target AoDs. Accordingly, the parameters of interest for target~$l$ are defined as
\begin{equation}
\left\{
\begin{aligned}
    &\boldsymbol{\hat{\Psi}}_{l,B}
    =
    \left[
        \hat{\tau}_{l,B},
        \hat{\mu}_{l,B}
    \right],\, \text{for the BS},  \\
    &\boldsymbol{\hat{\Psi}}_{l,m_r}
    =
    \left[
        \hat{\theta}_{l,B},
        \hat{\tau}_{l,m_r},
        \hat{\mu}_{l,m_r}
    \right],\, \text{for sensing Eve $m_r$}.
\end{aligned}
\right.
\end{equation}

For analytical tractability, the delay and Doppler shifts are assumed to lie on the discrete grid \cite{wei2025otfs}, i.e., $\tau_{l,i} = \frac{\ell_{l,i}}{N_c\Delta f}$ and $\mu_{l,i} = \frac{\nu_{l,i}}{N_sT_{\mathrm{tot}}},\, i\in\{B,m_r\}$, where $\ell_{l,i},\, \nu_{l,i}$ denote the corresponding delay and Doppler bins. Accordingly, the signals in \eqref{estimation-signal} and \eqref{Eve processed echo no AoA} can be expressed in matrix form as
\begin{equation}\label{Matrix form echo}
    \mathbf Y_i = \sum_{l\in\mathcal L} \tilde{\beta}_{l,i} \mathbf D_{\ell_{l,i}} \widetilde{\mathbf W}_l \mathbf D_{\nu_{l,i}} + \mathbf{N}_i,
\end{equation}
where
\begin{equation}
    \widetilde{\mathbf W}_l
    =
    \left(
        \mathbf I_{N_c}
        \otimes
        \mathbf a_t^{\mathrm H}(\theta_{l,B})
    \right)
    \begin{bmatrix}
        \overline{\mathbf W}_0\overline{\mathbf s}_0
        &
        \cdots
        &
        \overline{\mathbf W}_{N_s-1}\overline{\mathbf s}_{N_s-1}
    \end{bmatrix}, \nonumber
\label{eq:true_spatial_waveform}
\end{equation}
with
\begin{align}
    \overline{\mathbf W}_{n_s} &=
    \operatorname{blkdiag}
    \left(
        \mathbf W_{0,n_s},
        \ldots,
        \mathbf W_{N_c-1,n_s}
    \right)
    \in
    \mathbb C^{N_cN_t\times N_cK}, \notag \\
    \overline{\mathbf s}_{n_s} &=[
        \mathbf s_{0,n_s},
        \ldots,
        \mathbf s_{N_c-1,n_s}
    ]^\mathrm{H}
    \in
    \mathbb C^{N_cK\times1}.
\end{align}

Moreover,
\begin{align}
    \mathbf{D}_{\ell_{l,i}}
    &= \operatorname{diag}\left(
        1,e^{-j2\pi\ell_{l,i}\frac{1}{N_c}},\ldots,
        e^{-j2\pi\ell_{l,i}\frac{N_c-1}{N_c}}
      \right),  \label{delay mat}\\
    \mathbf{D}_{\nu_{l,i}}
    &= \operatorname{diag}\left(
        1,e^{j2\pi\nu_{l,i}\frac{1}{N_s}},\ldots,
        e^{j2\pi\nu_{l,i}\frac{N_s-1}{N_s}}
      \right). \label{doppler mat}
\end{align}

To perform matched filtering, we first declare the reference signal available at receiver $i$:
\begin{equation}\label{eq:receiver_reference}
    \mathbf{x}^{\mathrm{ref}}_{n_c,n_s,i}
    =
    \begin{cases}
        \mathbf{W}_{n_c,n_s} \mathbf{s}_{n_c,n_s}, & i = B,\\
        \hat{\mathbf{x}}_{n_c,n_s}, & i = m_r.
    \end{cases}
\end{equation}

For a candidate AoD $\hat{\theta}$, the reference signal is projected onto the corresponding transmit steering direction. The resulting matrix $\mathbf{X}^{\mathrm{ref}}_i \left( \hat{\theta} \right) \in
\mathbb{C}^{N_c\times N_s}$ is defined element-wise as
\begin{equation}\label{eq:candidate_reference}
    \left[
        \mathbf{X}^{\mathrm{ref}}_i \left( \hat{\theta} \right)
    \right]_{n_c,n_s}
    =
    \mathbf{a}_t^{\mathrm H}(\hat{\theta})
    \mathbf{x}^{\mathrm{ref}}_{n_c,n_s,i}.
\end{equation}

The MF is performed element-wise between \eqref{Matrix form echo} and \eqref{eq:candidate_reference}. The delay-Doppler map (DDM) is obtained by an IFFT across the subcarriers and an FFT across the symbols \cite{li2025sensing, wang2024multipath}:
\begin{equation}\label{eq:MF_expression}
    \mathbf{\Xi}_i \left( \hat{\theta} \right)
    =
    \mathbf{F}_{N_c}^{\mathrm H}
    \left(
        \mathbf{Y}_i
        \odot
        \left[
            \mathbf{X}^{\mathrm{ref}}_i \left( \hat{\theta} \right)
        \right]^*
    \right)
    \mathbf{F}_{N_s},
\end{equation}
where $\mathbf F_{N_c}$ and $\mathbf F_{N_s}$ denote the DFT matrices along the subcarrier and symbol dimensions, respectively.

\subsubsection{BS}
For the monostatic BS, the AoD is equivalent to the AoA $\theta_{l,B}$, and therefore $\mathbf\Xi_B(\theta_{l,B})$ directly yields the DDM. Then, a constant false alarm rate (CFAR) detector is applied to $\left|\mathbf\Xi_B(\theta_{l,B})\right|^2$ to identify $\boldsymbol{\hat{\Psi}}_{l,B}$.

\subsubsection{Sensing Eve}
For the bistatic sensing Eve $m_r$, target AoDs are unknown. Therefore,
\eqref{eq:MF_expression} is evaluated over an AoD search grid
$\Theta_t=\{\theta_1,\ldots,\theta_{N_\theta}\}$. Stacking the resulting DDMs gives the three-dimensional AoD-delay-Doppler map
$\mathcal X_{m_r} \in \mathbb C^{N_\theta\times N_c\times N_s}$, whose entries are defined as
$\mathcal X_{m_r}[m,\ell,\nu] = \left[ \mathbf{\Xi}_{m_r}(\theta_m) \right]_{\ell,\nu}$. A CFAR detector can be employed to $\left|\mathcal X_{m_r}[m,\ell,\nu] \right|^2$ to identify $\boldsymbol{\hat{\Psi}}_{l,m_r}$.

\begin{figure}[t]
    \centering
    \includegraphics[width=0.9\linewidth]{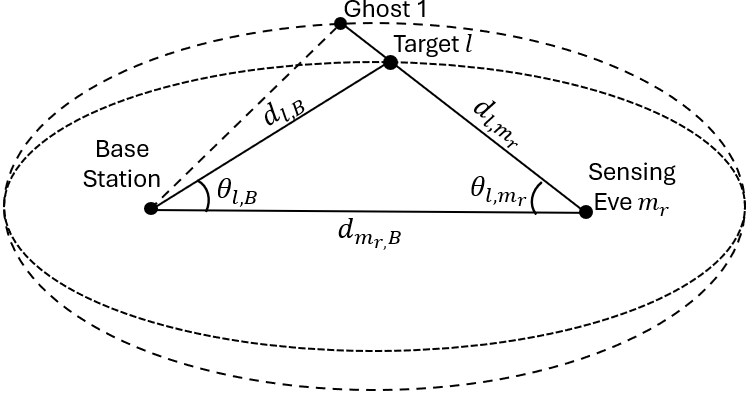}
    \caption{Illustration of the sensing security vulnerability in an ISAC system, where $d_{l,B}$ is the BS-target distance, $d_{m_r, B}$ is the BS-Eve distance, and $d_{l, m_r}$ is the target-Eve distance.}
    \label{fig:security issue}
\end{figure}

An important distinction is that the presence of an AG does not imply that it has the same impact on all receivers. The BS has prior knowledge of the true target AoD/AoA according to Assumption~\ref{assumption3} and therefore evaluates the delay-Doppler map along the known target direction. In the considered design, the ghost angular locations are selected to be distinct from the known target and user directions, allowing the BS to reject the corresponding ghost hypotheses during target estimation. In contrast, a sensing Eve does not know the target AoD and must search over the full candidate angle grid $\Theta_t$, so the generated AGs remain plausible target hypotheses. Hence, no additional AG suppression constraint is required at the BS under the assumed prior knowledge of the target angle.

\begin{remark}\label{remark 1}
    Under \textbf{Assumption~\ref{assumption1}}, the sensing Eve knows the location of the BS and, consequently, the BS-Eve distance $d_{m_r, B}$. We further consider a worst-case scenario in which the sensing Eve can accurately estimate the target AoA $\theta_{l,m_r}$. As illustrated in Fig.~\ref{fig:security issue}, the estimated AoA constrains the target to lie along a ray originating from the sensing Eve. If the bistatic delay $\tau_{l,m_r}$ is also recovered, the corresponding bistatic range $d_{l,B}+d_{l,m_r}=c\tau_{l,m_r}$ defines an ellipse with the BS and the sensing Eve as its foci. The intersection of the AoA ray with this ellipse can therefore determine the target position. Conversely, an inaccurate delay estimate produces an incorrect ellipse, which leads to an ambiguous target location. Alternatively, if the target AoD $\theta_{l,B}$ is recovered, it defines a second ray originating from the BS. The intersection of the AoD and AoA rays can then determine the target position without requiring knowledge of the bistatic delay. Hence, ambiguity or errors in the recovered AoD can likewise result in incorrect target localization. Therefore, given reliable knowledge of the target AoA, the recovery of either the bistatic delay or the target AoD can be sufficient for target localization. In addition, the Doppler shift $\mu_{l,m_r}$ conveys information about the target motion and can facilitate continuous target tracking. These observations motivate protecting the target AoD, delay, and Doppler information from potential sensing Eves.
\end{remark}

\subsection{Communication Models}

For a communication user $k\in\mathcal{K}$, following the standard OFDM signal processing procedures, the received frequency-domain signal is given by
\begin{equation}\label{Comm signal}
    \mathbf{y}_{n_c, n_s, k}
    = \mathbf{H}_{n_c,n_s,k}
      \mathbf{x}_{n_c,n_s}
      + {\mathbf{n}}_{n_c,n_s, k}, 
\end{equation}
where $\mathbf{H}_{n_c,n_s,k}\in \mathbb{C}^{N_k \times N_t}$ denotes the downlink channel vector on subcarrier $n_c$ and symbol $n_s$, and $\mathbf{n}_{n_c,n_s,k}~\sim~\mathcal{CN}(0, \sigma_c^2 \mathbf{I})$ is AWGN. Perfect CSI is assumed at the BS, obtainable through standard pilot estimation prior to the beamforming design. 

Substituting \eqref{Tx signal} into \eqref{Comm signal}, we can decompose the received signal as
\begin{align} \label{Comm signal decomp}
    \mathbf{y}_{n_c, n_s, k} &= \mathbf{H}_{n_c,n_s,k} \mathbf{w}_{n_c,n_s,k} s_{n_c,n_s,k} + \mathbf{H}_{n_c,n_s,k} \mathbf{q}_{n_c,n_s} \\
    & \; + \sum_{k'\in\mathcal{K}, k'\neq k} \mathbf{H}_{n_c,n_s,k} \mathbf{w}_{n_c,n_s,k'} s_{n_c,n_s,k'} + {\mathbf{n}}_{n_c,n_s,k}, \notag
\end{align}
where $\mathbf{w}_{n_c,n_s,k} \in \mathbb{C}^{N_t \times 1} $ is the $k$-th column of $\mathbf{W}_{n_c, n_s}$.

For a hypothetical communication Eve $m_c \in \mathcal{M}_c$, $m_c \notin \mathcal{K} \cup \mathcal{L}$ located at angle $\theta_{m_c} \in \left[-\frac{\pi}{2}, \frac{\pi}{2}\right]$ , the intercepted signal in relation to the $k$-th CU is given by
\begin{align} \label{Comm Eve signal}
    & {y}_{n_c, n_s, k, m_c} = \mathbf{h}_{n_c,n_s,m_c} \mathbf{w}_{n_c,n_s,k} s_{n_c,n_s,k} + \mathbf{h}_{n_c,n_s,m_c} \mathbf{q}_{n_c,n_s} \notag \\
    &\,+ \sum_{k'\in\mathcal{K}, k'\neq k} \mathbf{h}_{n_c,n_s,m_c} \mathbf{w}_{n_c,n_s,k'} s_{n_c,n_s,k'} + n_{n_c,n_s,m_c},
\end{align}
where $\mathbf{h}_{n_c,n_s,m_c} = \mathbf{a}_{t}^{\mathrm{H}}(\theta_{m_c})$ denotes the downlink channel vector for the communication Eve, and $n_{n_c,n_s,m_c}$ is AWGN.

We note that the channel coefficient and the receive steering vector are omitted in \eqref{Comm Eve signal}, because the BS has no knowledge of Eve's distance, as stated in Assumption~\ref{assumption2}, or the number of communication Eve's receive antennas. Hence, the channel toward the communication Eve is characterized only in the hypothetical transmit angular domain rather than modeled as a full MIMO channel.

\section{Performance Metrics}

In this section, we present the performance metrics for sensing, communication, and security, respectively.

\subsection{Sensing Quality at the BS}

To characterize the sensing capability of the BS, we adopt the sensing SNR as the performance metric \cite{demirhan2024cell}. A higher sensing SNR generally leads to improved target detection and parameter estimation performance. Since the BS has perfect knowledge of the transmitted reference signal and the AN is nulled toward the target directions, the sensing SNR for target $l$ at each $(n_c,n_s)$ resource element 
is given by:
\begin{equation}\label{eq:BS_SNR}
\gamma_{n_c, n_s, l, B}
=
\frac{ \sigma_{\beta}^2
       \left\| \mathbf{a}_{t}^{\mathrm{H}}(\theta_{l,B}) \mathbf{W}_{n_c,n_s}
       \right\|^{2}
     }{\sigma_r^{2}},
\end{equation}
where $\sigma_{\beta}^2$ denotes the average power of the channel coefficient.

\subsection{Sensing Security via Ambiguity Function Shaping}

The MF processing in \eqref{eq:MF_expression} reveals the AoD-delay-Doppler ambiguity characteristics induced by the beamforming matrices. To analytically characterize these ambiguity properties, we adopt the normalized AF \cite{han2025sensing} and assume sensing Eve's reference signal is error-free, i.e., $\hat{\mathbf{x}}_{n_c,n_s} = \mathbf{x}_{n_c,n_s}$.

For target $l$, let $\hat{\theta} \in\!\left[-\tfrac{\pi}{2},\tfrac{\pi}{2}\right]$ denote a discrete candidate AoD, and define the delay and Doppler error as $\Delta\ell = \hat{\ell}-\ell$, $\Delta\nu = \hat{\nu}-\nu$, respectively. By neglecting the channel coefficient and AWGN, the normalized AoD-delay-Doppler AF is defined in \eqref{AF}.
\begin{table*}
\centering
\begin{minipage}{1\textwidth}
    \begin{align}\label{AF}
    &\chi
    \left(
        \Delta\ell,
        \Delta\nu,
        \hat{\theta},
        \theta_{l,B}
    \right) =
    \frac{1}{N_cN_s}
    \mathbb E
    \Bigg[
    \sum_{n_c\in\mathcal N_c}
    \sum_{n_s\in\mathcal N_s}
    \left(
        \mathbf W_{n_c,n_s}\mathbf s_{n_c,n_s}
        +
        \mathbf q_{n_c,n_s}
    \right)^{\mathrm H}
    \mathbf a_t(\hat{\theta})
    \mathbf a_t^{\mathrm H}(\theta_{l,B})
    \mathbf W_{n_c,n_s}\mathbf s_{n_c,n_s} 
    e^{j2\pi n_c\Delta\ell/N_c}
    e^{-j2\pi n_s\Delta\nu/N_s}
    \Bigg].
\end{align}
\medskip
\hrule
\end{minipage}
\end{table*}
With $\mathbb E[\mathbf s_{n_c,n_s}\mathbf s_{n_c,n_s}^{\mathrm H}]= \mathbf I$ and $\mathbb E[ \mathbf s_{n_c,n_s}\mathbf q_{n_c,n_s}^{\mathrm H}] =\mathbf 0$, which is specified in \eqref{Tx signal}, the cross term associated with the AN vanishes in expectation. Hence, \eqref{AF} reduces to
\begin{align}\label{AF_compact}
    &\chi
    \left(
        \Delta\ell,
        \Delta\nu,
        \hat{\theta},
        \theta_{l,B}
    \right)
    \nonumber\\
    &=
    \frac{1}{N_cN_s}
    \sum_{n_c\in\mathcal N_c}
    \sum_{n_s\in\mathcal N_s}
    e^{j2\pi n_c\Delta\ell/N_c}
    e^{-j2\pi n_s\Delta\nu/N_s}
    \nonumber\\
    &\qquad\times
    \operatorname{Tr}
    \left(
        \mathbf a_t(\hat{\theta})
        \mathbf a_t^{\mathrm H}(\theta_{l,B})
        \mathbf W_{n_c,n_s}
        \mathbf W_{n_c,n_s}^{\mathrm H}
    \right).
\end{align}

For notational convenience, define
\begin{align*}
    & \overline{\mathbf W}_{n_c,n_s} =
    \mathbf W_{n_c,n_s}
    \mathbf W_{n_c,n_s}^{\mathrm H}
    \in\mathbb C^{N_t\times N_t}, \\
    & \overline{\mathbf{W}} = \mathrm{blkdiag} \left( \overline{\mathbf{W}}_{1,1}, \ldots, \overline{\mathbf{W}}_{N_c,N_s} \right) \in\mathbb C^{N_t N_c N_s \times N_t N_c N_s}, \\
    &\bar{\mathbf{A}}(\theta_{l,B}) = \frac{1}{\sqrt{N_c N_s}} \mathbf{I}_{N_cN_s} \otimes \mathbf{a}_t(\theta_{l,B}).
\end{align*}

Accordingly, the AF can be equivalently written as
\begin{align}\label{AF_trace}
    &\chi
    \left(
        \Delta\ell,
        \Delta\nu,
        \hat{\theta},
        \theta_{l,B}
    \right) \notag \\
    & \; = \mathrm{Tr}\left(\bar{\mathbf{A}}(\hat{\theta})
       (\mathbf{D}_{\Delta \ell}\otimes\mathbf{D}_{\Delta \nu})
       \bar{\mathbf{A}}^{\mathrm{H}}(\theta_{l,B}) \overline{\mathbf{W}} \right),
\end{align}
where $\mathbf{D}_{\Delta \ell}$, $\mathbf{D}_{\Delta \nu}$ has been defined in \eqref{delay mat} and \eqref{doppler mat}.

For the sensing Eves, strong sidelobes at prescribed ghost profiles are generated to mislead their target estimations. Since the same beamforming matrix determines the AF observed by all receivers, the AF must be carefully shaped to simultaneously preserve the sensing performance at the BS and generate AGs at the sensing Eves. To quantify and control such AF shaping, we adopt the peak-to-sidelobe level (PSL) and integrated sidelobe level (ISL) 
Specifically, the PSL characterizes the largest sidelobe peak relative to the mainlobe 
whereas the ISL measures the aggregate sidelobe energy relative to the mainlobe \cite{li2024dual, li2024mimo} 
The PSL is formulated as
\begin{equation}\label{PSL}
    \mathrm{PSL}_{l} = \frac{
        \left|\chi(\Delta\ell_g,\Delta\nu_g,\theta_{g}, \theta_{l,B})\right|^2
    }{
        \left|\chi(0,0,\theta_{l,B},\theta_{l,B})\right|^2
    }, \forall g \in \mathcal{G}, g\notin\mathcal{L} \cup \mathcal{K},
\end{equation}
where $\mathcal{G}$ denotes the set of ghost indices, 
and each ghost has the AoD-delay-Doppler profile of $\left(\theta_{g}, \Delta\ell_g, \Delta\nu_g \right)$. 

The ISL formulation is given by:
\begin{equation}\label{ISL}
    \mathrm{ISL}_{l} =\frac{
     \sum_{\hat{\theta}} \sum_{
    \Delta \ell = 0,\;\Delta \nu = 0}^{N_c-1,\;N_s-1}
    \left|
    \chi(\Delta \ell,\Delta \nu,\hat{\theta}, \theta_{l,B})
    \right|^2}{ \left|\chi(0,0,\theta_{l,B},\theta_{l,B})\right|^2
    } -1 ,
\end{equation}
where $-1$ is to subtract the mainlobe from the summation.

\subsection{Communication and Sensing Securities Using Artificial Noise}

The sensing SINR at a sensing Eve is a fundamental metric for characterizing its sensing performance. By comparing its SINR with the sensing SNR achieved at the BS, the sensing security of the system can be assessed \cite{zhou2025beamforming}. 
However, evaluating sensing Eve's sensing SINR requires knowledge of the statistical characteristics of its reference and surveillance channels, which are unavailable at the BS. Therefore, the sensing SINR at the sensing Eve cannot be directly employed.

Since AN is introduced to contaminate the reference signal observed by a potential sensing Eve, we characterize the quality of Eve's recovered reference signal using its achievable data rate \cite{musallam2025enhancing}. A lower achievable rate indicates a higher level of AN in the reference signal, thereby impairing the sensing Eve's MF processing and degrading its AoD-delay-Doppler estimation accuracy.

\begin{remark}
It is worth emphasizing that the achievable rate metric here is not intended to represent an information-theoretic communication rate. It serves, rather, as a tractable surrogate quantifying the quality of the reference signal available to the sensing Eve when the BS has no instantaneous knowledge of the sensing Eve's reference and surveillance channels. Its relationship with the sensing Eve's actual sensing SINR is numerically evaluated in Section~\ref{sec:simulation}.
\end{remark}

For a hypothetical sensing Eve located in the angular direction $\theta_{m_r}$, neither its channel coefficient nor the number of receive antennas is known at the BS. Therefore, we consider the transmission rate associated with the signal power radiated toward the direction $\theta_{m_r}$. The resulting data rate is given by \eqref{Eve ref rate}. In this equation, $\mathbf{W}_{n_c, n_s, k} = \mathbf{w}_{n_c, n_s, k} \mathbf{w}_{n_c, n_s, k}^{\mathrm{H}}$ and $\mathbf{Q}_{n_c, n_s} = \mathbf{q}_{n_c, n_s} \mathbf{q}_{n_c, n_s}^{\mathrm{H}}$.

\begin{remark}\label{remark 2}
    We note that the channel coefficient is omitted in \eqref{Eve ref rate} due to the lack of knowledge about Eve's location (i.e., Assumption~\ref{assumption2}). As a result, \eqref{Eve ref rate} characterizes the best-case sensing capability of Eve. In practice, propagation loss would further attenuate the received signal and hence reduce the performance for sensing Eves \cite{musallam2025enhancing, niu2025survey}.
\end{remark}

\begin{table*}[t]
\centering
\begin{minipage}{\textwidth}
\begin{align}
& C_{m_r} = \frac{1}{N_c N_s} \sum_{n_c \in \mathcal{N}_c, n_s \in \mathcal{N}_s} \log_2 \left( 1 + \frac{\mathbf{a}_{t}^{\mathrm{H}}\left(\theta_{m_r}\right) \sum_{k \in \mathcal{K}} \mathbf{W}_{n_c, n_s, k}  \mathbf{a}_{t}\left(\theta_{m_r}\right)}{ \mathbf{a}_{t}^{\mathrm{H}}\left(\theta_{m_r}\right) \mathbf{Q}_{n_c,n_s} \mathbf{a}_{t}\left(\theta_{m_r}\right) + \sigma_{c}^{2} }  \right), \label{Eve ref rate}\\
&  C_{k, m_c} = \frac{1}{N_c N_s} \sum_{n_c \in \mathcal{N}_c, n_s \in \mathcal{N}_s} \log_2 \left( 1 + \frac{\mathbf{a}_{t}^{\mathrm{H}}(\theta_{m_c}) \mathbf{W}_{n_c,n_s, k} \mathbf{a}_{t}(\theta_{m_c})}{ \mathbf{a}_{t}^{\mathrm{H}}(\theta_{m_c}) \left(\sum_{k' \in \mathcal{K}, k' \neq k} \mathbf{W}_{n_c,n_s, k'} + \mathbf{Q}_{n_c,n_s}\right) \mathbf{a}_{t}(\theta_{m_c}) + \sigma_{c}^{2}} \right), \label{comm Eve data rate}
\end{align}
\medskip
\hrule
\end{minipage}
\end{table*}

On the other hand, for a hypothetical communication Eve $m_c$, its data rate regarding user $k$ is shown in \eqref{comm Eve data rate}. 

\subsection{Communication Performance}

Based on the communication signal model in \eqref{Comm signal decomp}, the achievable rate of CU $k$ on resource element $(n_c,n_s)$ is denoted by $C_{n_c,n_s,k}$ and equals the logarithmic term in \eqref{CU data rate}. Its average over all resource elements is $C_k$ in \eqref{CU data rate}. The optimization imposes the QoS threshold on every $C_{n_c,n_s,k}$, rather than only on the average $C_k$.

\begin{table*}[t]
\centering
\begin{minipage}{\textwidth}
\begin{align}
C_k
=
\frac{1}{N_cN_s}
\sum_{\substack{n_c\in\mathcal N_c\\ n_s\in\mathcal N_s}}
\log_2\det
\Bigg(
\mathbf I
+
\mathbf H_{n_c,n_s,k}
\mathbf W_{n_c,n_s,k}
\mathbf H_{n_c,n_s,k}^{\mathrm H}
\Bigg(
\mathbf H_{n_c,n_s,k}
\Big(
\sum_{k'\in\mathcal K, k'\neq k}
\mathbf W_{n_c,n_s,k'}
+
\mathbf Q_{n_c,n_s}
\Big)
\mathbf H_{n_c,n_s,k}^{\mathrm H}
+
\sigma_c^2\mathbf I
\Bigg)^{-1}
\Bigg).
\label{CU data rate}
\end{align}
\medskip
\hrule
\end{minipage}
\end{table*}

\begin{remark}\label{remark 5}
    In the considered system, neither the locations nor the CSI of the communication Eves are available at the BS. Therefore, an explicit secrecy rate \cite{aman2026integrating}, defined as $\min_{k \in \mathcal{K}} \left[C_{k} - \max_{m_c \in \mathcal{M}_c} C_{k,m_c}\right]^{+}$, cannot be adopted. Instead, communication security is ensured by guaranteeing the achievable rates of the legitimate CUs while constraining the achievable rates of potential Eves.
\end{remark}

\section{Dual-Secure Design for MIMO-OFDM ISAC Systems}

This section formulates the dual-secure ISAC design problem and proposes an algorithm for solving the resulting non-convex problem.

\subsection{Problem Formulation} 

The design objective is to ensure reliable sensing performance while simultaneously guaranteeing communication performance, communication security, and sensing security. Since the locations of potential sensing and communication Eves are unknown to the BS, their angular domains are discretized into candidate grids
\[ \Theta_{\rm E}^{r} = \{\theta_{1}^{r},\ldots,\theta_{M_r}^{r}\}, \qquad \Theta_{\rm E}^{c} = \{\theta_{1}^{c},\ldots,\theta_{M_c}^{c}\},\]
where each grid point represents a hypothetical Eve direction. The security constraints are enforced over all candidate directions, thereby avoiding the need for knowledge of the actual Eve locations or CSI. 

Accordingly, the optimization problem is formulated based on the aforementioned performance metrics:
\begin{subequations}\label{opt 1}
\begin{align}
&\max_{\{\mathbf{W}_{n_c,n_s,k}\succeq 0\},\,\{\mathbf{Q}_{n_c,n_s}\succeq 0\},\,\eta_B}\quad \eta_B \label{opt1 obj}\\
&\text{s.t.}\;\; \gamma_{n_c,n_s,l,B}\geq\eta_B,\; \forall n_c\in\mathcal{N}_c,\ \forall n_s\in\mathcal{N}_s,\ \forall l\in\mathcal{L}, \label{opt1 b}\\
&\qquad C_{m_r}\leq\eta_E,\; \forall m_r\in\mathcal{M}_r, \label{opt1 c}\\
&\qquad C_{k,m_c}\leq\eta_E,\; \forall k\in\mathcal{K},\ \forall m_c\in\mathcal{M}_c, \label{opt1 c2}\\
&\qquad \mathrm{PSL}_{l}\geq\eta_P,\; \forall l\in\mathcal{L}, \label{opt1 d}\\
&\qquad \mathrm{ISL}_{l}\geq\eta_I,\; \forall l\in\mathcal{L}, \label{opt1 e}\\
&\qquad C_{n_c,n_s,k}\geq\eta_S,\;\forall n_c\in\mathcal{N}_c,\ \forall n_s\in\mathcal{N}_s,\ \forall k\in\mathcal{K}, \label{opt1 f}\\
&\qquad \mathbf{a}_t^{H}(\theta_{l,B})\mathbf{Q}_{n_c,n_s}\mathbf{a}_t(\theta_{l,B})=0, \nonumber\\
&\hspace{3cm} \forall n_c\in\mathcal{N}_c,\ \forall n_s\in\mathcal{N}_s,\ \forall l\in\mathcal{L}, \label{opt1 null}\\
&\qquad \sum_{n_c\in\mathcal{N}_c}\sum_{n_s\in\mathcal{N}_s}\!\!\mathrm{Tr}\!\left(\sum_{k\in\mathcal{K}}\mathbf{W}_{n_c,n_s,k}+\mathbf{Q}_{n_c,n_s}\right)\leq P_t, \label{opt1 g}\\
&\qquad \mathrm{rank}\!\left(\mathbf{W}_{n_c,n_s,k}\right)=1,\; \forall n_c\in\mathcal{N}_c,\ \forall n_s\in\mathcal{N}_s,\ \forall k\in\mathcal{K}, \label{opt1 i}\\
&\qquad \mathrm{rank}\!\left(\mathbf{Q}_{n_c,n_s}\right)=1,\; \forall n_c\in\mathcal{N}_c,\ \forall n_s\in\mathcal{N}_s. \label{opt1 iq}
\end{align}
\end{subequations}

The objective in \eqref{opt1 obj}, together with constraint \eqref{opt1 b}, maximizes the minimum sensing SNR across all targets and OFDM resource elements. This max-min formulation prevents the sensing performance from being dominated by a subset of favorable targets or resource elements and guarantees a common sensing SNR level $\eta_B$.
Constraints \eqref{opt1 c} and \eqref{opt1 c2} respectively limit the reference signal quality at sensing Eves and the intercepted user rates at communication Eves. Constraint \eqref{opt1 d} enforces the prescribed ghost peak strength, whereas the lower bound in \eqref{opt1 e} requires aggregate sidelobe energy outside the true target mainlobe; increasing $\eta_I$ raises this sidelobe energy requirement, which can make individual ghosts and targets less detectable under CFAR due to the elevated noise floor. Constraint \eqref{opt1 f} ensures the quality of service (QoS) for each communication user on every resource element. The AN nulling constraint ensures that no artificial noise power is radiated toward the known target directions, thereby preserving the legitimate sensing echoes and remaining consistent with the signal models in Sections~II-B and II-C. Constraint \eqref{opt1 g} enforces the transmit power budget $P_{t}$. The rank-one conditions \eqref{opt1 i} and \eqref{opt1 iq} express the single vector beamforming model. We drop these non-convex conditions when solving the semidefinite relaxation. After the iterative optimization converges, Gaussian randomization is applied to the optimized positive-semidefinite matrices to obtain rank-one transmit beamformers and AN vectors.

\subsection{Problem Transformation}

The optimization problem is non-convex due to constraints \eqref{opt1 c}-\eqref{opt1 f}. 
Each constraint is expressed in difference-of-convex (DC) form and approximated using a first-order Taylor expansion, yielding the convex surrogates in \eqref{opt1 c transform} and \eqref{opt1 c transform2}.
Here, we define
\begin{equation*}
\left \{
\begin{aligned}
    &\mathbf{T}_{n_c, n_s} =  \sum_{k \in \mathcal{K}} \mathbf{W}_{n_c, n_s, k} + \mathbf{Q}_{n_c, n_s},\\[-2pt]
    &\mathbf{T}^{(e)}_{n_c, n_s} = \sum_{k \in \mathcal{K}} \mathbf{W}_{n_c, n_s, k}^{(e)} + \mathbf{Q}_{n_c, n_s}^{(e)}, \\[-2pt]
    & \mathbf{T}_{n_c, n_s, -k} = \sum_{k' \in \mathcal{K}, k' \neq k} \mathbf{W}_{n_c,n_s, k'}+ \mathbf{Q}_{n_c,n_s},\\
    & \mathbf{T}_{n_c, n_s, -k}^{(e)} = \sum_{k' \in \mathcal{K}, k' \neq k} \mathbf{W}_{n_c,n_s, k'}^{(e)} + \mathbf{Q}_{n_c,n_s}^{(e)},
\end{aligned}
\right.
\end{equation*}
where the matrices ${\mathbf{W}}_{n_c,n_s,k}^{(e)}$ and $\mathbf{Q}_{n_c,n_s}^{(e)}$ are updated during each iteration.

\begin{table*}[t]
\centering
\begin{minipage}{\textwidth}
\begin{align}
&\frac{1}{N_c N_s}
\sum_{\substack{n_c \in \mathcal{N}_c,\\ n_s \in \mathcal{N}_s}} \log \left( \mathbf{a}_t^{\mathrm{H}}(\theta_{m_r}) \mathbf{T}^{(e)}_{n_c,n_s} \mathbf{a}_t(\theta_{m_r}) +\sigma_c^2 \right) -
\log \left( \mathbf{a}_t^{\mathrm H}(\theta_{m_r}) \mathbf{Q}_{n_c,n_s} \mathbf{a}_t(\theta_{m_r}) +\sigma_c^2 \right) 
+ \frac{ \mathbf{a}_t^{\mathrm H}(\theta_{m_r}) \left( \mathbf{T}_{n_c,n_s} - \mathbf{T}^{(e)}_{n_c,n_s} \right) \mathbf{a}_t(\theta_{m_r})}{ \mathbf{a}_t^{\mathrm H}(\theta_{m_r}) \mathbf{T}^{(e)}_{n_c,n_s} \mathbf{a}_t(\theta_{m_r}) +\sigma_c^2}\leq \eta_E \ln 2, \label{opt1 c transform} \\
&\frac{1}{N_c N_s}\!\!\sum_{\substack{n_c \in \mathcal{N}_c,\\ n_s \in \mathcal{N}_s}}\!\!\!\!\log\left( \mathbf{a}_t^{\mathrm H}(\theta_{m_c}) \mathbf{T}^{(e)}_{n_c,n_s} \mathbf{a}_t(\theta_{m_c}) +\sigma_c^2 \right) - \log \left(
\mathbf{a}_t^{\mathrm H}(\theta_{m_c}) \mathbf{T}_{n_c, n_s, -k} \mathbf{a}_t(\theta_{m_c}) +\sigma_c^2 \right) + \frac{ \mathbf{a}_t^{\mathrm H}(\theta_{m_c}) \left( \mathbf{T}_{n_c,n_s} - \mathbf{T}^{(e)}_{n_c,n_s}\right) \mathbf{a}_t(\theta_{m_c})}{\mathbf{a}_t^{\mathrm H}(\theta_{m_c}) \mathbf{T}^{(e)}_{n_c,n_s} \mathbf{a}_t(\theta_{m_c}) +\sigma_c^2 } \leq \eta_E \ln 2, \label{opt1 c transform2}
\end{align}
\medskip
\hrule
\end{minipage}
\end{table*}

Using the same approach, constraints \eqref{opt1 d} and \eqref{opt1 e} are replaced by the convex surrogate constraints in \eqref{opt1 d transform} and \eqref{opt1 e transform}, respectively. In those two equations, we define
\begin{equation*}
\left\{
\begin{aligned}
    f_{\mathrm{P}} \left(\overline{\mathbf{W}}^{(e)}\right)
    &\triangleq
    \mathrm{Tr}\!\left(
    \bar{\mathbf{A}}(\theta_{g})
    (\mathbf{D}_{\Delta \ell_g}\otimes\mathbf{D}_{\Delta \nu_g})
    \bar{\mathbf{A}}^{\mathrm{H}}(\theta_{l,B})
    \overline{\mathbf{W}}^{(e)}
    \right),\\
    f_{\mathrm{I}}\!\left(\overline{\mathbf{W}}^{(e)}\right)
    &\triangleq
    \mathrm{Tr}\!\left(
    \bar{\mathbf{A}}(\hat{\theta})
    (\mathbf{D}_{\Delta \ell}\otimes\mathbf{D}_{\Delta \nu})
    \bar{\mathbf{A}}^{\mathrm{H}}(\theta_{l,B})
    \overline{\mathbf{W}}^{(e)}
    \right).
\end{aligned}
\right.
\end{equation*}
and the auxiliary matrix $\overline{\mathbf W}^{(e)}$ is constructed from $\mathbf W_{n_c,n_s,k}^{(e)}$.

Finally, following the same procedure, constraint \eqref{opt1 f} is approximated by the convex surrogate shown in \eqref{opt1 f transform 2}.

\begin{table*}[t]
\centering
\begin{minipage}{\textwidth}
\begin{align}
&\eta_{{P}}
\Big|
    \mathrm{Tr} \left(
        \bar{\mathbf{A}}(\theta_{l,B})
        \bar{\mathbf{A}}^{\mathrm{H}}(\theta_{l,B})
        \overline{\mathbf{W}}
    \right)
\Big|^{2}
-
\Bigg(
    \left| f_{\mathrm{P}}\left(\overline{\mathbf{W}}^{(e)}\right) \right|^{2}
     +
    2 \Re \bigg\{
        f_{\mathrm{P}}^{*}(\overline{\mathbf{W}}^{(e)}) \Big[
            \mathrm{Tr}\left(
                \bar{\mathbf{A}}(\theta_{g})
                (\mathbf{D}_{\ell_{g}}\otimes\mathbf{D}_{\nu_{g}})
                \bar{\mathbf{A}}^{\mathrm{H}}(\theta_{l,B})
                \left(\overline{\mathbf{W}}
                - \overline{\mathbf{W}}^{(e)} \right)
            \right)
        \Big]
    \bigg\}
\Bigg) \leq 0, \label{opt1 d transform} \\
& \left( \eta_{{I}} + 1 \right)
\Big| \mathrm{Tr} \left(
        \bar{\mathbf{A}}(\theta_{l,B})
        \bar{\mathbf{A}}^{\mathrm{H}}(\theta_{l,B})
        \overline{\mathbf{W}}
    \right)
\Big|^{2}
- \!\!\!\!\sum_{\hat{\theta}, \Delta \ell, \Delta \nu} \Bigg(
    \left| f_{\mathrm{I}}\left(\overline{\mathbf{W}}^{(e)}\right) \right|^{2}
     +
    2 \Re \bigg\{
        f_{\mathrm{I}}^{*}(\overline{\mathbf{W}}^{(e)}) \Big[
            \mathrm{Tr}\left(
                \bar{\mathbf{A}}(\hat{\theta})
                (\mathbf{D}_{\Delta \ell}\otimes\mathbf{D}_{\Delta \nu})
                \bar{\mathbf{A}}^{\mathrm{H}}(\theta_{l,B})
                \left(\overline{\mathbf{W}}
                - \overline{\mathbf{W}}^{(e)} \right)
            \right)
        \Big]
    \bigg\}
\Bigg) \leq 0, \label{opt1 e transform}
\end{align}
\medskip
\hrule
\end{minipage}
\end{table*}

\begin{table*}[t]
\centering
\begin{minipage}{\textwidth}
\begin{align}\label{opt1 f transform 2}
& \log \det \left( \mathbf{H}_{n_c,n_s,k} \mathbf{T}_{n_c, n_s} \mathbf{H}_{n_c,n_s,k}^{\mathrm{H}} + \sigma_c^2 \mathbf{I} \right) - \log \det \left( \mathbf{H}_{n_c,n_s,k} \mathbf{T}^{(e)}_{n_c, n_s, -k} \mathbf{H}_{n_c,n_s,k}^{\mathrm{H}} + \sigma_c^2 \mathbf{I} \right) \notag\\
&- \mathrm{Tr} \left( \left( \mathbf{H}_{n_c,n_s,k} \mathbf{T}^{(e)}_{n_c, n_s, -k} \mathbf{H}_{n_c,n_s,k}^{\mathrm{H}} + \sigma_c^2 \mathbf{I} \right)^{-1} \left( \mathbf{H}_{n_c,n_s,k} \left( \mathbf{T}_{n_c, n_s, -k} -\mathbf{T}_{n_c, n_s, -k}^{(e)} \right) \mathbf{H}_{n_c,n_s,k}^{\mathrm{H}} \right) \right) \geq \eta_{S}\ln 2,\; \forall k\in\mathcal K,\ \forall n_c\in\mathcal N_c,\ \forall n_s\in\mathcal N_s.
\end{align}
\medskip
\hrule
\end{minipage}
\end{table*}

Therefore, at iteration $e$, problem~\eqref{opt 1} is approximated by the following convex surrogate subproblem:
\begin{subequations} \label{opt1 convex}
\begin{align}
    &\max_{\{\mathbf{W}_{n_c,n_s,k}\succeq 0\},\,\{\mathbf{Q}_{n_c,n_s}\succeq 0\},\,\eta_B}\quad \eta_B \\
    & \text{s.t.}\;\; \gamma_{n_c,n_s,l,B} \geq \eta_B, \;
    \forall n_c\in\mathcal N_c,\;
    \forall n_s\in\mathcal N_s,\;
    \forall l\in\mathcal L,\\
    &\hspace{0.6cm}\eqref{opt1 null}, \;\eqref{opt1 g}, \;
    \eqref{opt1 c transform},\;
    \eqref{opt1 c transform2},\;
    \eqref{opt1 d transform},\;
    \eqref{opt1 e transform},\;
    \eqref{opt1 f transform 2}.
\end{align}
\label{opt 1 convex}
\end{subequations}


For fixed expansion points at iteration $e$, the objective is linear in $\eta_B$ and all surrogate constraints are convex. Hence, problem~\eqref{opt 1 convex} constitutes a convex subproblem that can be efficiently solved using standard convex optimization solvers.
The iterative procedure for solving problem~\eqref{opt 1 convex} is shown in Algorithm~\ref{alg1}. Each iteration optimizes $(K+1)N_cN_s$ Hermitian matrices of order $N_t$; this is the variable count and does not characterize the SDP solver runtime. The runtime also depends on the number of angular grids, AF shaping constraints (i.e., the number of ghosts), and on the solver. The convergence condition is defined as $\left\|{\mathbf{W}}^{(e)}_{n_c, n_s, k} - {\mathbf{W}}^{(e-1)}_{n_c, n_s, k}\right\| \leq \epsilon$ and $\left\| \mathbf{Q}^{(e)}_{n_c, n_s} - \mathbf{Q}^{(e-1)}_{n_c, n_s}\right\| \leq \epsilon$.

\begin{algorithm}[h]
\caption{Iterative Optimization for Solving Problem~\eqref{opt1 convex}}\label{alg1}
\begin{algorithmic}[1]
\STATE Set the iteration number $e = 1$, and randomly initialize $\mathbf{W}_{n_c, n_s, k}^{\left(0\right)}$ and $\mathbf{Q}_{n_c, n_s}^{\left(0\right)}$.
\REPEAT
    \STATE Solve \eqref{opt 1 convex} to obtain $\mathbf{W}_{n_c, n_s, k}$ and $\mathbf{Q}_{n_c, n_s}$.
    \STATE Update $\mathbf{W}_{n_c, n_s, k}^{\left(e\right)}$, $\mathbf{Q}_{n_c, n_s}^{\left(e\right)}$, and $\overline{\mathbf{W}}^{(e)}$.
    \STATE $e = e + 1$.
\UNTIL{converge.}
\normalsize
\STATE Apply Gaussian randomization to the optimized $\mathbf{W}_{n_c,n_s,k}$ and $\mathbf{Q}_{n_c,n_s}$ to obtain rank-one beamformers $\mathbf{w}_{n_c,n_s,k}$ and AN vectors $\mathbf{q}_{n_c,n_s}$.
\end{algorithmic}
\end{algorithm}

\section{Simulation Results}\label{sec:simulation}

\begin{table}[!t]
\centering
\caption{Simulation Settings}
\label{tab:Sim parameters}
\renewcommand{\arraystretch}{2.5}
\begin{tabular}{|c|c|c|}
\hline
Parameter & Symbol & Value \\
\hline
\hline
Carrier frequency & $f_{c}$ & 24~GHz\\
\hline
OFDM symbol duration & $T$ & $8.33 \times 10^{-6}$ \\
\hline
Cyclic prefix duration & $T_{\mathrm{cp}}$ & $0.59 \times 10^{-6}$ \\
\hline
Number of subcarriers & $N_{c}$ & 16 \\
\hline
Number of symbols & $N_{s}$ & 10 \\
\hline
Number of transmit/receive antennas & $N_{t}, N_{r} $ & 16 \\
\hline
CU's receive antennas & $N_{k}$ & 2 \\
\hline
Sensing Eve's receive antennas & $N_{m_r}$ & 24 \\
\hline
Noise power & $\sigma_{c}^2,\,\sigma_{r}^2$ & -90~dBm \\
\hline
Power budget & $P_{t}$ & $15~\text{dBm}$ \\
\hline
\end{tabular}
\end{table}

This section presents numerical results to evaluate the performance of the proposed dual-secure ISAC framework. Unless otherwise specified, all simulation parameters are set to the default values listed in Table~\ref{tab:Sim parameters}. The BS and all receivers employ uniform linear arrays (ULAs) with half-wavelength antenna spacing, and quadrature phase-shift keying (QPSK) modulation is adopted. The angular domain is discretized in $1^\circ$ increments over the range $-90^\circ$ to $90^\circ$.
In the considered two-dimensional Cartesian coordinate system, the BS is located at the origin, while a static sensing target is located at $(12,11)$, resulting in both the delay and Doppler bins being zero, $\ell_{l} = \nu_{l} =0$. 
Two legitimate CUs are positioned at $(6,-8)$ and $(18,-2)$, respectively. The sensing Eve and communication Eve are randomly located within the region $x\in[0,65]$ and $y\in[-65,65]$ for evaluation; their sampled locations are not provided to the beamforming optimizer. The ghost profiles, represented by $(\theta_g,\Delta\ell_g,\Delta\nu_g)$, are set to $(-42,5,7)$ and $(12,3,4)$. All channels are modeled using geometry-based channel models. Finally, the convergence tolerance in Algorithm~\ref{alg1} is set to $\epsilon=10^{-3}$.

Three benchmark schemes are considered for comparison. The first is a non-secure baseline obtained by removing both the AN-based security constraints and the AG-generation constraints, consistent with conventional ISAC beamforming designs such as \cite{luo2023ris}. The second is an AN-only baseline, obtained by removing the AG-related PSL and ISL constraints while retaining the AN-based sensing security constraints. This benchmark isolates the protection provided by AG and is motivated by AN-based sensing security designs such as \cite{musallam2025enhancing}. The third is an AG-only baseline, in which AN-based security constraints are removed while the AF shaping constraints for AG-generation are retained, motivated by secure AF shaping designs such as \cite{du2025securing}. These benchmarks isolate the contributions of AN and/or AG and enable comparison with the proposed joint design.

\subsection{Sensing versus Communication}

\begin{figure}[!t]
    \centering
    \includegraphics[width=0.9\linewidth]{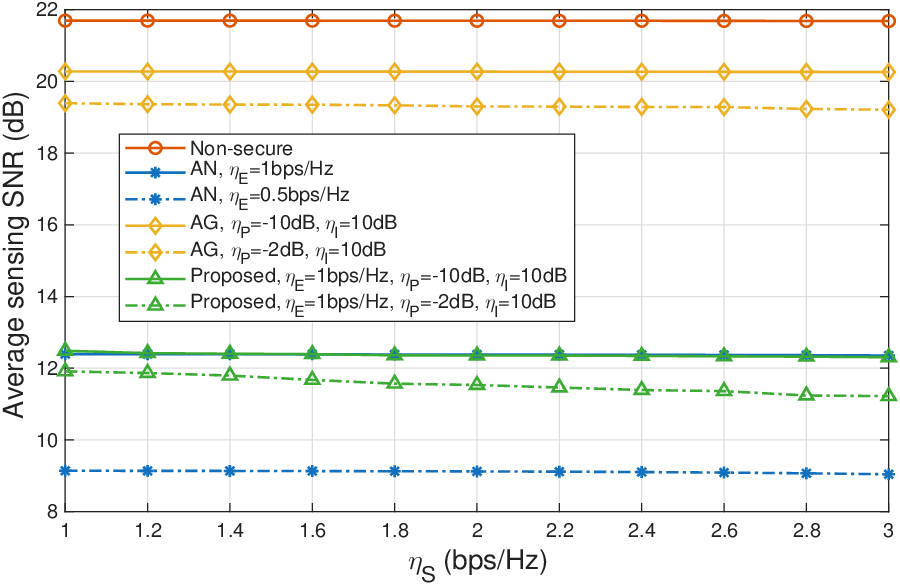}
    \caption{Average sensing SNR of the BS versus the QoS requirement $\eta_{S}$.}
    \label{fig:sensingSNR_etaS}
\end{figure}

We first investigate the trade-off between sensing performance, characterized by the sensing SNR in \eqref{eq:BS_SNR}, and communication performance, characterized by the QoS requirement $\eta_S$ in \eqref{opt1 f transform 2}. As shown in Fig.~\ref{fig:sensingSNR_etaS}, the non-secure design achieves the highest sensing SNR over the entire range of $\eta_S$, with only a slight performance degradation as the QoS requirement increases. For the AN-only scheme, when $\eta_E=1$ bps/Hz, the sensing SNR is substantially reduced compared with the non-secure design, and a more stringent $\eta_E$ leads to a further reduction. This sensing performance degradation is the cost of ensuring communication security via AN, 
as part of the transmit power must be allocated to AN rather than target illumination. In contrast, for the AG-only scheme, its sensing SNR is slightly lower than that of the non-secure design but remains considerably higher than that of the AN-only scheme, since all transmit power is utilized for target illumination while simultaneously shaping the desired AF pattern. 
Moreover, a more stringent AG requirement, i.e., increasing $\eta_P$ from -10dB to -2dB, leads to a reduction in sensing SNR, as stronger ghost generation requires more aggressive AF shaping, thereby limiting the achievable gain at the true target.

For the proposed scheme, communication and sensing security are simultaneously enforced. Consequently, the sensing SNR is lower than that of all benchmarks, reflecting the additional performance cost of jointly satisfying the two security objectives. When the required ghost strength is relatively low, e.g., $\eta_{P}=-10$ dB, the sensing SNR achieved by the proposed scheme remains close to that of the AN-only scheme under the same $\eta_E$. This indicates that a less stringent sensing security requirement 
incurs only a minor loss in sensing performance. However, as stronger ghosts are required, the sensing SNR is further reduced. In this case, the sensing performance degradation also becomes more pronounced as the communication QoS requirement $\eta_S$ increases. This result reveals that simultaneously ensuring communication performance, sensing security, and communication security comes at the cost of degraded sensing performance.

\subsection{Communication Security via Artificial Noise}

\begin{figure}[!t]
    \centering
    \includegraphics[width=0.9\linewidth]{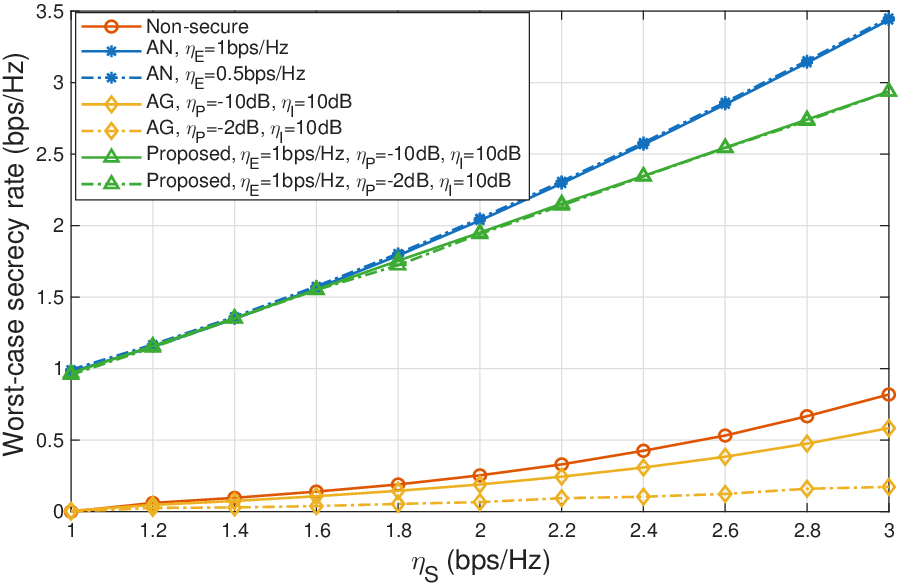}
    \caption{Worst-case secrecy rate versus the QoS requirement~$\eta_{S}$. }
    \label{fig:SR_etaS}
\end{figure}

To evaluate communication security, we calculate the worst-case secrecy rate as described in Remark \ref{remark 5}. 
As shown in Fig.~\ref{fig:SR_etaS}, the non-secure and AG-only schemes achieve lower secrecy rates than the AN-only and proposed schemes because neither explicitly incorporates communication security constraints. 
Increasing the QoS requirement provides only a minor communication security improvement for these schemes, as the higher QoS requirement is accompanied by increased information leakage to the communication Eve.

In contrast, the AN-only and proposed dual-secure schemes achieve substantially higher secrecy rates because the achievable rate of the communication Eve is explicitly constrained. Moreover, their secrecy rates increase significantly with $\eta_S$, since a more stringent QoS requirement improves the achievable rates of the legitimate users while the communication Eve's rate remains bounded by $\eta_E$. A smaller $\eta_E$ further enhances the secrecy rate, although the corresponding gain is relatively limited. At relatively low $\eta_S$, the secrecy rate of the proposed scheme remains close to that of the AN-only scheme, whereas the gap gradually widens as $\eta_S$ increases. This gap reflects the cost of simultaneously maintaining sensing security and communication performance, namely, a reduction in communication security.

\subsection{Sensing Security via Artificial Noise}

\begin{figure}[!t]
    \centering
    \includegraphics[width=0.9\linewidth]{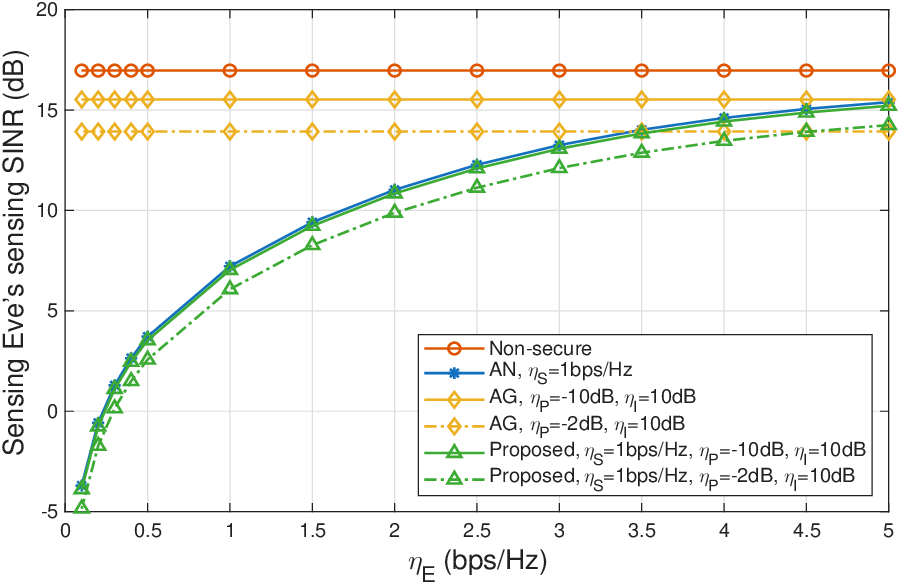}
    \caption{Sensing Eve's sensing SINR versus data rate upper bound $\eta_{E}$.}
    \label{fig:sensingEveSINR_etaE}
\end{figure}

To demonstrate the effectiveness of AN in enhancing sensing security and to verify Remark~\ref{remark 2}, we evaluate the sensing Eve's sensing SINR \cite{zhao2022joint} versus the upper bound on the achievable rate of the reference signal (i.e., $\eta_E$). A lower sensing SINR at the sensing Eve leads to reduced parameter estimation accuracy and therefore indicates stronger sensing security. 

As shown in Fig.~\ref{fig:sensingEveSINR_etaE}, the sensing SINR of the non-secure and AG-only schemes remains unchanged with $\eta_E$, since neither scheme employs AN to distort the reference signal. For the AG-only scheme, a larger $\eta_P$ slightly reduces the sensing SINR, as stronger ghosts introduce more interference, thereby degrading the sensing quality at the sensing Eve. In contrast, the sensing SINRs of the AN-only and proposed schemes decrease significantly as $\eta_E$ becomes more stringent. A smaller $\eta_E$ creates stronger AN in the sensing Eve's reference signal, which introduces greater mismatch and impairment during MF 
and consequently reduces the sensing SINR. Moreover, the proposed scheme achieves a lower sensing SINR as $\eta_P$ increases, indicating that creating stronger ghosts introduces additional interference in the echo signal and further enhances sensing security. Fig.~\ref{fig:SR_etaS} and Fig.~\ref{fig:sensingEveSINR_etaE} demonstrate the capability of the proposed scheme in simultaneously enhancing communication and sensing security. 

Although Fig.~\ref{fig:sensingEveSINR_etaE} demonstrates that reducing the data rate of the sensing Eve's reference signal can effectively decrease its sensing SINR and thereby enhance sensing security, a sensing Eve equipped with a large antenna array may exploit its increased spatial degrees of freedom (DoFs) to partially suppress the AN through spatial processing. This may allow the sensing Eve to preserve a relatively high-quality reference signal, thereby leaving sensing privacy concerns unresolved. To address this vulnerability, the proposed scheme introduces a second layer of sensing security through ghost target generation, whose effectiveness is investigated in the following subsection.

\subsection{Sensing Security via Artificial Ghosts}

\subsubsection{One Dimensional}

\begin{figure}[!t]
    \centering
    \begin{subfigure}{0.45\textwidth}
        \centering
        \includegraphics[width=0.9\linewidth]{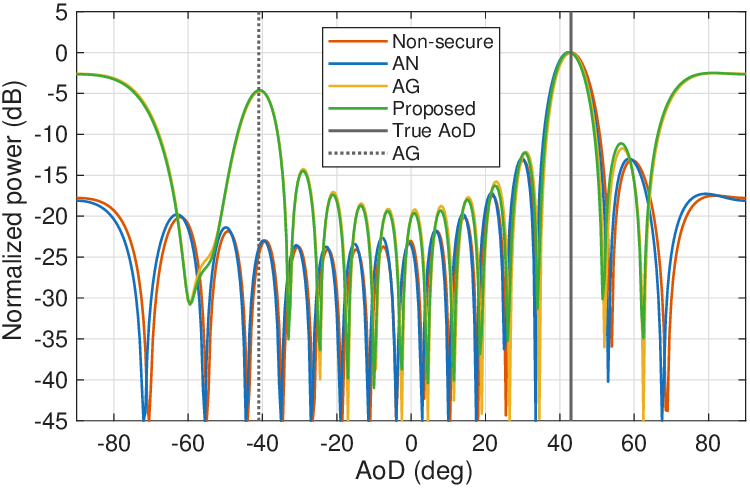}
        \caption{Comparison with benchmark schemes under $\eta_{P}=-5$ dB and $\eta_{I}=15$ dB.}
    \end{subfigure}
    \begin{subfigure}{0.45\textwidth}
        \centering
        \includegraphics[width=0.9\linewidth]{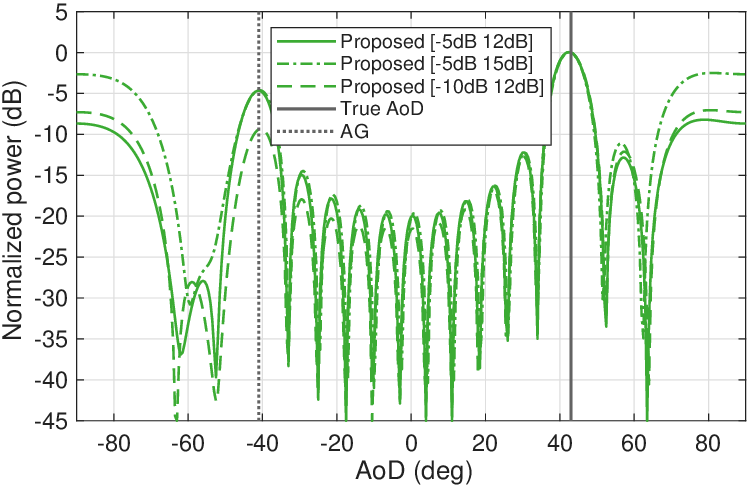}
        \caption{Comparison under different $[\eta_{P} \; \eta_{I}]$. }
    \end{subfigure}
    \caption{The AoD map obtained from the zero delay, zero Doppler slice of the AoD–delay–Doppler map, the parameters are $\eta_{S}=1$ bps/Hz and $\eta_{E}=3$ bps/Hz.}
    \label{fig:AODmap}
\end{figure}

We first demonstrate ghosts in the angular domain by examining the corresponding zero delay-Doppler slice of the AoD–delay–Doppler map. For this purpose, a single ghost is introduced at $\theta_g=-42^{\circ}$ with $\Delta\ell_g=\Delta\nu_g=0$. The non-secure and AN-only schemes are not designed to generate AGs, therefore, only one dominant peak is observed at the true target AoD of $43^{\circ}$, as shown in Fig.~\ref{fig:AODmap}(a). In contrast, the proposed scheme exhibits two distinct peaks corresponding to the true and ghost AoDs, respectively, closely matching the AG-only benchmark. This result demonstrates that the proposed design can generate angular ghosts to enhance sensing security while maintaining communication performance and security. When an estimation algorithm such as MUSIC \cite{zhang2010direction} is employed by the sensing Eve, both the true and ghost targets may be identified as candidate targets, thereby introducing ambiguity into the sensing Eve's target estimation.

To further illustrate the effects of the PSL and ISL on ghost generation, we evaluate the proposed scheme on the same AoD map under varying PSL and ISL requirements. As shown in Fig.~\ref{fig:AODmap}(b), for $\eta_{P}=-5$ dB, increasing $\eta_{I}$ results in a higher overall sidelobe level. Consequently, the ghost peak may become less distinguishable from the surrounding sidelobes and may no longer be reliably identified as a candidate target. For a fixed $\eta_{I}$, increasing $\eta_{P}$ strengthens the ghost peak, thereby facilitating its identification by the sensing Eve. These results indicate that $\eta_{P}$ and $\eta_{I}$ should be carefully selected to ensure that the generated AGs remain detectable by the sensing Eve. Although not shown here, similar conclusions can be drawn for other one-dimensional AGs.

\subsubsection{Two Dimensional}

\begin{figure}[!t]
    \centering
    \begin{subfigure}{0.45\textwidth}
    \centering
    \includegraphics[width=0.9\linewidth]{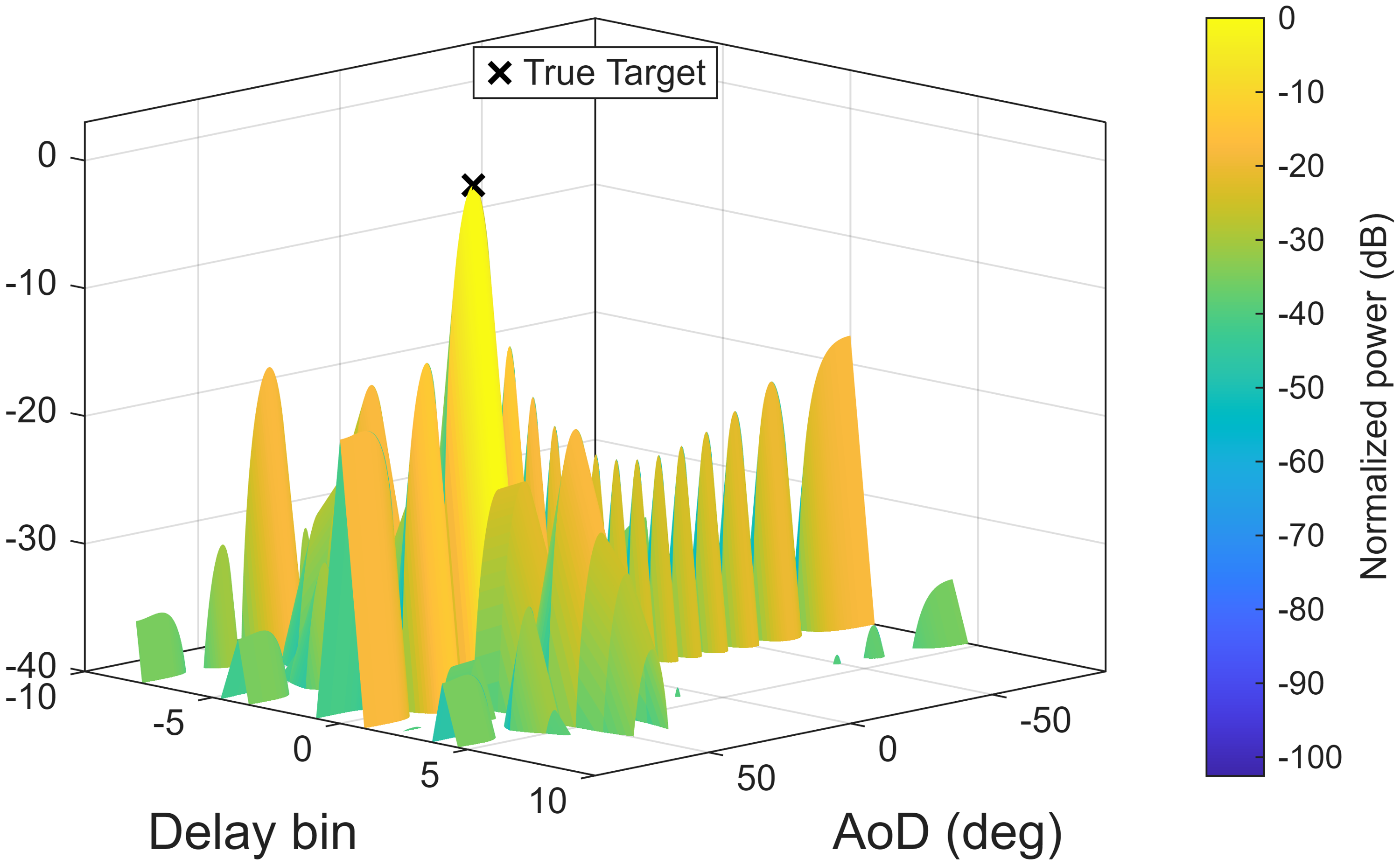}
    \caption{Sensing Eve AoD-delay map without security.}
    \vspace{5pt}
    \end{subfigure}
    \begin{subfigure}{0.45\textwidth}
    \centering
    \includegraphics[width=0.9\linewidth]{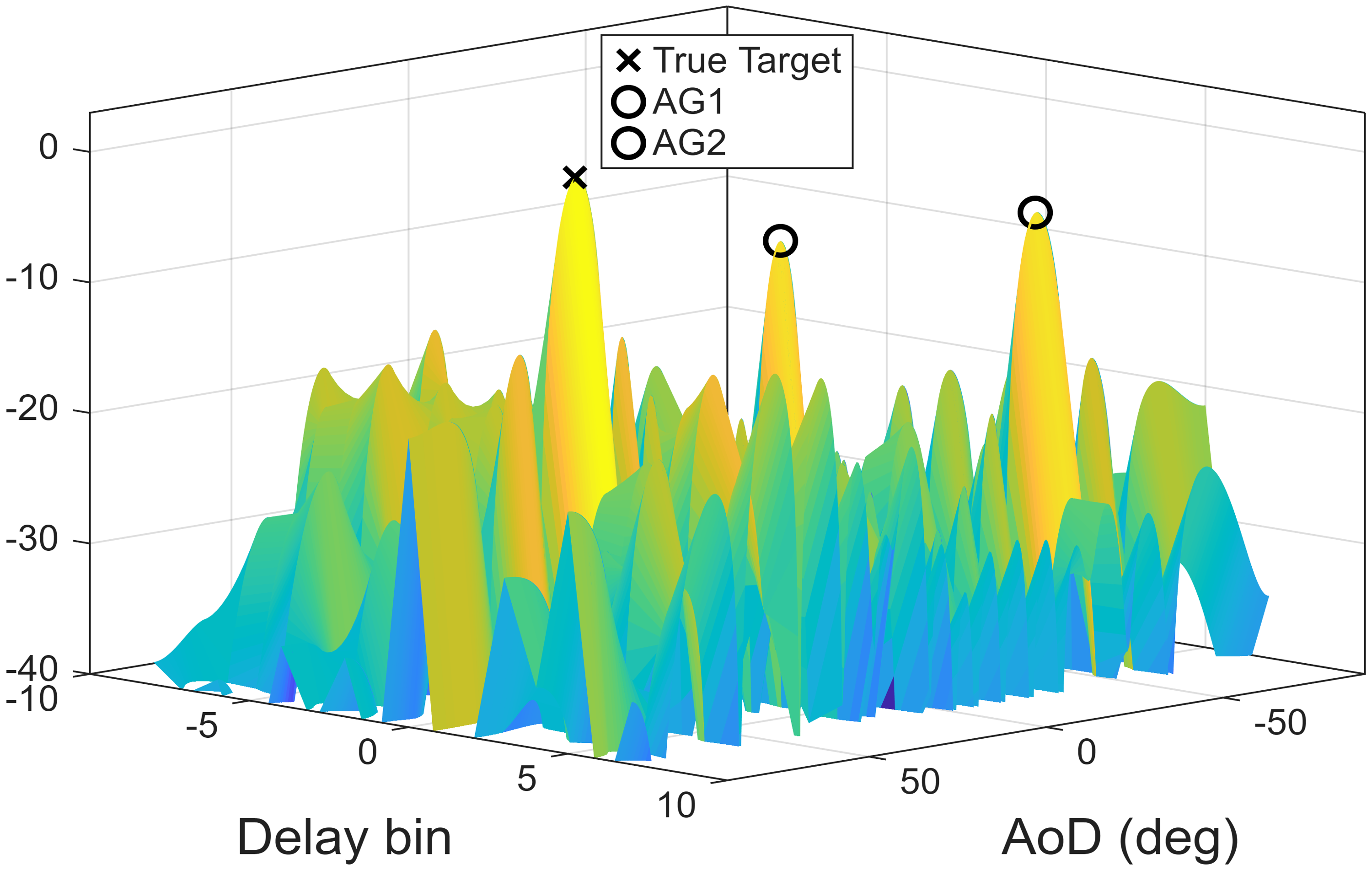}
    \caption{Sensing Eve AoD-delay map with security.}
    \vspace{5pt}
    \end{subfigure}
    \begin{subfigure}{0.45\textwidth}
    \centering
    \includegraphics[width=0.8\linewidth]{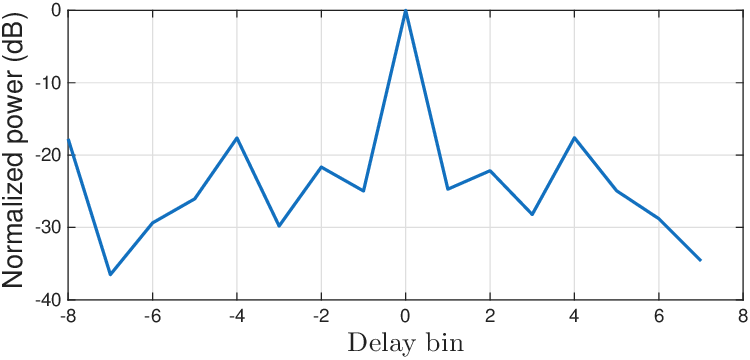}
    \caption{BS delay map.}
    \end{subfigure}
    \caption{AoD-delay and delay maps for the sensing Eve and the BS, respectively. The parameters are $\eta_{P}=-5$ dB, $\eta_{I}=12$ dB, $\eta_{S}=1$ bps/Hz and $\eta_{E}=3$ bps/Hz.}
    \label{fig:2D_maps}
\end{figure}

We next demonstrate two-dimensional AGs under the zero Doppler slice, corresponding to secure sensing of a static target, as discussed in Remark~\ref{remark 1}. For this purpose, we set the AG profiles to $(-42^\circ,5,0)$ and $(12^\circ,3,0)$. For the sensing Eve, the resulting two-dimensional AoD-delay maps are shown in Fig.~\ref{fig:2D_maps}(a) and (b). As observed in Fig.~\ref{fig:2D_maps}(a), the non-secure design yields a single dominant peak associated with the true target, allowing the sensing Eve to readily identify its profile and thus posing a sensing security risk. In contrast, as shown in Fig.~\ref{fig:2D_maps}(b), the proposed design generates three distinct peaks: the true target at an AoD of $43^\circ$ and zero delay, and two ghosts located at $-42^\circ$ and $12^\circ$, with delay indices of $5$ and $3$, respectively. Consequently, when a CFAR detector is applied at the sensing Eve, all three peaks can be detected as candidate targets, although only one corresponds to the actual target. This ambiguity prevents the sensing Eve from reliably identifying the true target.

For the BS, since the true target AoD is known, we examine the delay profile obtained from the slice corresponding to the true AoD and zero Doppler. As shown in Fig.~\ref{fig:2D_maps}(c), a single dominant peak appears at the true delay, with no AGs present. This confirms that the proposed design preserves the BS's sensing performance while introducing ambiguities for the sensing Eve. Similar observations apply to the other two-dimensional ghost configurations, namely AoD-Doppler and delay-Doppler ghosts.

\subsubsection{Three Dimensional}

\begin{figure}[t]
    \centering
    \begin{subfigure}{0.45\textwidth}
    \centering
    \includegraphics[width=0.9\linewidth]{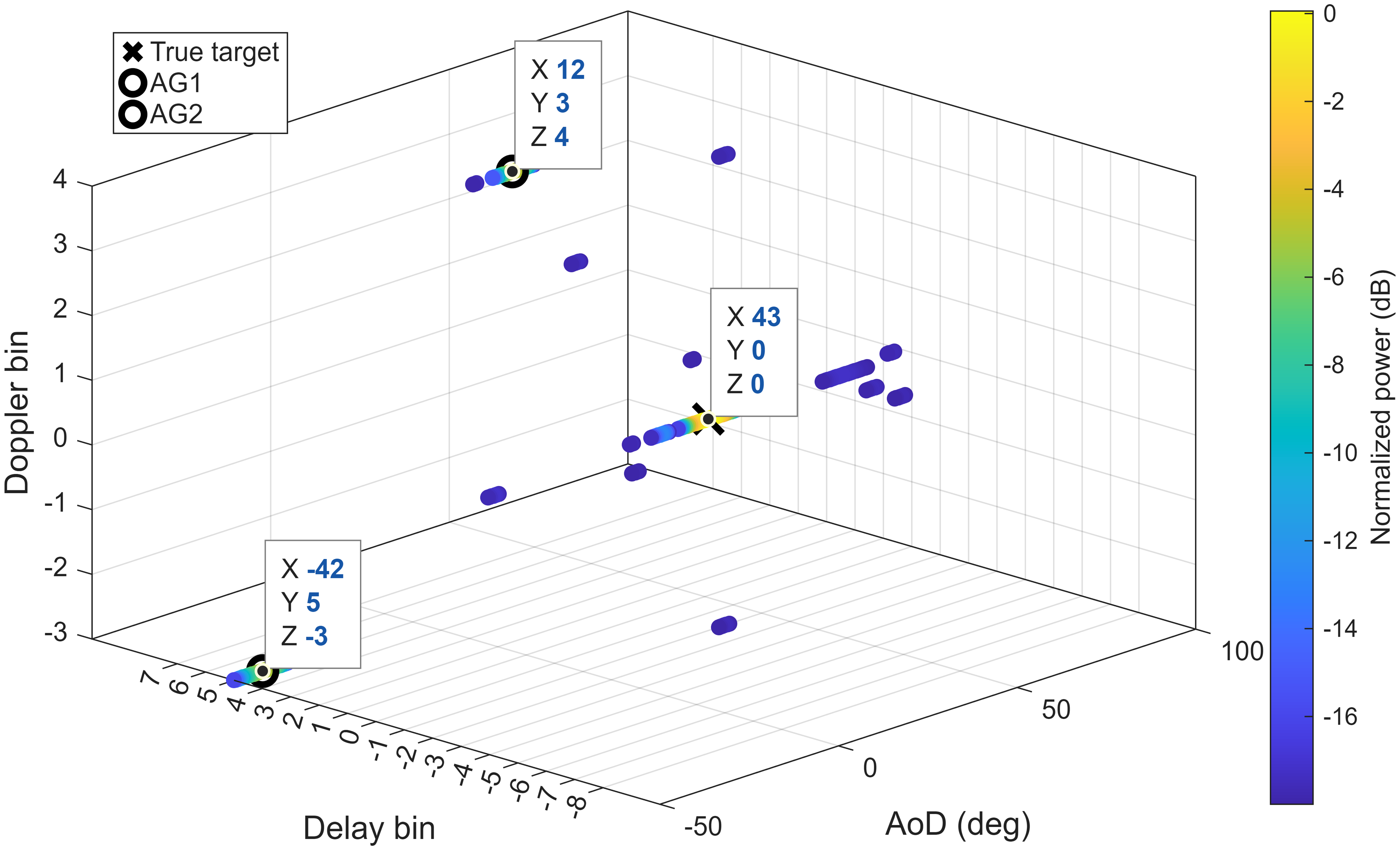}
    \caption{Sensing Eve AoD-delay-Doppler map.}
    \vspace{5pt}
    \end{subfigure}
    \begin{subfigure}{0.45\textwidth}
    \centering
    \includegraphics[width=0.8\linewidth]{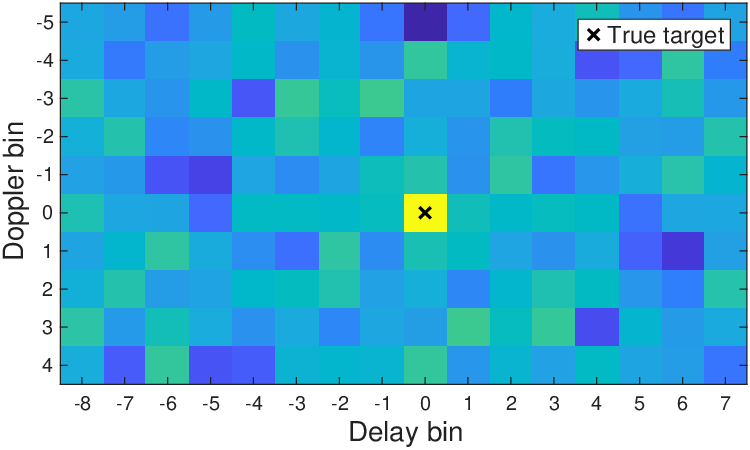}
    \caption{BS delay-Doppler map.}
    \end{subfigure}
    \caption{AoD-delay-Doppler and delay-Doppler maps for the sensing Eve and the BS, respectively. The parameters are $\eta_{P}=-5$ dB, $\eta_{I}=12$ dB, $\eta_{S}=1$ bps/Hz and $\eta_{E}=3$ bps/Hz.}
    \label{fig:3D map}
\end{figure}

We finally examine the three-dimensional AoD-delay-Doppler map observed by the sensing Eve to illustrate the generated AGs. For clarity of visualization, only values above $-20$ dB are displayed. As shown in Fig.~\ref{fig:3D map}(a), the true target produces the strongest peak, while the two generated AGs appear at their prescribed locations with relative powers of approximately $-5$ dB, as intended by the PSL requirement, i.e., $\eta_{P}=-5$ dB. Note that, because the Doppler axis is displayed using centered positive and negative indices after \texttt{fftshift}, the wrapped Doppler bin $7$ is shown as bin $-3$. Consequently, when a target estimation algorithm is applied, the sensing Eve may identify all three peaks as candidate targets, thereby introducing ambiguity into the estimation results and enhancing sensing security.

In Fig.~\ref{fig:3D map}(b), we present the BS delay-Doppler map evaluated at the true AoD. A single dominant peak is observed at the true target location, with no AGs present. This observation is consistent with previous results and further confirms that the proposed design enhances sensing security by introducing AGs at the sensing Eve while preserving the sensing performance of the BS.

\section{Conclusion}


In this paper, we propose a CSI-agnostic dual-secure framework for mitigating both sensing and communication eavesdropping in MIMO-OFDM ISAC systems. 
Signal models were established for the BS, legitimate communication users, sensing Eves, and communication Eves. 
To enhance both communication and sensing security, AN was incorporated into the transmit beamforming design while being nulled toward the known sensing target directions.
The AN degrades the received signals at communication Eves and impairs the sensing capability of sensing Eves, thereby limiting their communication and sensing performance. However, capable sensing Eves may cancel the AN, allowing them to retain strong sensing capability. To address this case, an additional sensing security mechanism was jointly designed by intentionally generating AGs with fake AoD-delay-Doppler profiles. These AGs can be suppressed only by the legitimate sensing receiver, thereby reducing the probability that the sensing Eves correctly identify the true targets. To realize the proposed dual-secure design, the transmit beamformers were optimized to maximize the sensing SNR at the BS while satisfying multiple communication, sensing, and security constraints. These include a minimum achievable rate for legitimate users, maximum allowable achievable rates at potential sensing and communication Eves, and PSL and ISL constraints for AG generation. Numerical results demonstrated the effectiveness of the proposed framework and revealed the inherent trade-offs among communication performance, communication security, sensing performance, and sensing security.

\bibliographystyle{IEEEtran}
\bibliography{Ref}
\end{document}